\documentstyle[prb,aps,preprint,tighten]{revtex}
\begin{document}

\title{Variational theory of elastic manifolds with
correlated disorder and localization of interacting
quantum particles}

\author{Thierry Giamarchi}
\address{Laboratoire de Physique des Solides, Universit{\'e} Paris--Sud,
                   B{\^a}t. 510, 91405 Orsay, France\cite{junk}}
\author{Pierre Le Doussal}
\address{CNRS-Laboratoire de Physique Th{\'e}orique
de l'Ecole Normale Sup{\'e}rieure \cite{frad}, 24 Rue Lhomond, F-75231 Paris
Cedex,
France}
\maketitle

\begin{abstract}
We apply the gaussian variational method (GVM) to
study the equilibrium statistical mechanics of
the two related systems:
(i) classical elastic manifolds, such as flux lattices,
in presence of columnar disorder correlated along the $\tau$ direction
(ii) interacting quantum particles in a static random potential.
We find localization by disorder, the localized phase being described
by a replica symmetry broken solution confined to the mode $\omega=0$.
For classical systems we compute the correlation function
of relative displacements. In $d=2+1$, in the absence of dislocations,
the GVM allows to describes the Bose glass phase.
Along the columns the displacements saturate at a length $l_{\perp}$
indicating flux-line localization. Perpendicularly to the columns long range
order is destroyed. We find divergent tilt modulus $c_{44}=\infty$
and a $x \sim \tau^{1/2}$ scaling. Quantum systems
are studied using the analytic continuation
from imaginary to real time $\tau \to i t$.
We compute the conductivity and find that it behaves
at small frequency as $\sigma(\omega) \approx \omega^2$ in
all dimensions ($d < 4$) for which disorder is relevant.
We compute the quantum localization length $\xi$.
In $d=1$, where the model also describes interacting
fermions in a static random potential, we find a
delocalization transition and obtain analytically
both the low and high frequency behavior
of the conductivity for any value of the interaction.
We show that the marginality condition appears as the
condition to obtain the
correct physical behavior. Agreement with
renormalization group results
is found whenever it can be compared.

\end{abstract}

\section{Introduction}

Two related longstanding problems of statistical mechanics
have recently received renewed attention. On the one hand
the problem of the pinning of the flux lattice in high-Tc
superconductors has motivated further studies of the effect of
disorder on an elastic manifold
\cite{blatter_vortex_review}. Pinning by {\it correlated} disorder, e.g.
in the form of columnar defects introduced in the sample
by heavy-ion irradiation
\cite{civale_columnar_first,konczykowski_columnar_first},
is particularly interesting for technological applications
\cite{nelson_columnar_long,hwa_splay_prl,ledoussal_nelson_splay}
since it leads to a large increase in critical
current.
On the other hand, the problem of interacting quantum particles
in a random potential, as the one
of {\it quantum} fluctuations of an elastic manifold
in a random potential, have been recently studied
by a number of authors. The latter problems have many
experimental realizations, such as Charge Density Waves (CDW)
\cite{gruner_revue_cdw},
Wigner crystals \cite{rusin_shklovskii_wigner,normand_millis_wigner},
quantum vortex creep in
superconductors \cite{blatter_vortex_review}, superconducting-insulating
transition
in disordered superconductors
\cite{sorensen_bosons_disorder}.
There is a deep analogy between these two classes of
problems which rests on the well known identification of the bosons worldlines
in the imaginary-time path integral of quantum mechanics with actual flux lines
\cite{feynman_hibbs}. This relation
has been reanalyzed
\cite{nelson_seung_long,nelson_ledoussal_liquid,boson_mapping,%
kamien_ledoussal_nelson,chen_teitel_prl1,li_teitel_prb1,blatter_vortex_review}
in details and has
led for instance, to the prediction of a
new physical state of vortex systems \cite{nelson_columnar_long}
with vanishing linear resistivity,
where 3d flux lines are localized by columnar disorder, very much
like 2d quantum bosons become localized in presence of disorder
and form the so-called Bose glass state
\cite{fisher_bosons_scaling}.
Although there is some evidence for a Bose glass, or localized
phase, few analytical methods are available to study this
problem. One has to resort to series expansions,
numerical simulations
\cite{wallin_girvin_bosons,krauth_bosons_disorder,%
makivic_trivedi_bosons,batrouni_hwa}. Experimental evidence is also
controversial.
In the case of  $d=1$ (one spatial dimension, one
``time'' dimension) where interacting bosons and fermions can be
related, the problem can be studied using powerful
bosonization and Renormalization Group (RG) techniques.
These methods show \cite{giamarchi_loc} a transition away from
the superfluid phase when the repulsion is increased and allow to
compute the crossover towards the localized phase. However they
do not allow to describe the low energy properties of the resulting
(presumably) localized phase, since RG trajectories
runaway to strong coupling. Only in the case of non-interacting
fermions, or equivalently
bosons interacting with hard-core repulsion, is the ``Bose glass'' phase
rigorously known to exist. The conductivity was calculated
in that case by Berezinskii
\cite{berezinskii_conductivity_log} and found to
behave as $\sigma(\omega) \sim \omega^2 \ln(\omega)^2$.

An interesting method to study problems with an interplay of
elasticity and disorder
is the Gaussian Variational method (GVM) introduced by Mezard and
Parisi \cite{mezard_variational_replica}
to study the problem of elastic manifolds,
such as interfaces, in a random
{\it uncorrelated} potential.
In this method one finds the best quadratic Hamiltonian
approximation to the original complicated Hamiltonian. This
quadratic approximation is performed in replica space
and the optimization naturally leads to Replica Symmetry Breaking
(RSB) whenever the energy landscape is complicated and contains
metastable states. Within the gaussian ansatz, RSB appears as
the natural way to encode the distribution of the metastable
states induced by disorder. The possible optimal
Hamiltonian is parametrized by a function $\sigma(u)$, which leads to a large
number of optimization parameters. Thus one can hope
that it produces a reasonable approximation of the real problem.
Indeed, in the
interface problem, the method leads to a reasonable, Flory-like,
approximation for the roughness exponent; It also leads
to the correct physics of
sample to sample fluctuations \cite{hwa_fisher_flux}
in response functions, due to rare events. This was recently demonstrated in
$d=1+1$
using the mapping onto the
Burgers equation \cite{hwa_fisher_long}.
In recent papers \cite{giamarchi_vortex_short,giamarchi_vortex_long},
we have applied the GVM
to the problem of an elastic lattice in a random uncorrelated
potential. It was shown
\cite{giamarchi_vortex_short,giamarchi_vortex_long,korshunov_variational_short}
that the periodicity of the lattice leads to
logarithmic disorder-induced displacements
\cite{nattermann_pinning,villain_cosine_realrg}
at large scales in $d \ge 2$.
Comparison with two RG calculations
\cite{giamarchi_vortex_short,giamarchi_vortex_long,ledoussal_rsb_prl}
suggests that the GVM
captures the correct physics, and leads to reasonable quantitative
approximations. Although the GVM captures important non perturbative
features such as the existence of many local equilibrium states,
it may miss some details of the nonlinearities
and instanton-type configurations (kinks)
between metastable states.

In the present paper we apply the Gaussian Variational
Method to classical
problems with correlated disorder and
to quantum problems with disorder.
A summary of the results of this work has appeared
in \onlinecite{giamarchi_columnar_short}.
There are some additional subtleties in applying the
method, due
to translational invariance in one direction ( the ``imaginary time''
direction) and extra care should be paid in treating
boundary conditions and taking
the large ``imaginary time'' limit $\beta \rightarrow \infty$ properly.
We then find that the GVM naturally leads to localization
by disorder and we compute the properties in the localized phase.
For the classical system we obtain the tilt modulus and find that
$c_{44}(q_z) \to \infty$ as $q_z \to 0$ in the localized
phase, i.e a signature of the transverse Meissner effect
predicted in Ref. \onlinecite{nelson_columnar_long}.
For quantum systems we investigate in detail dynamical correlation
functions such as the conductivity.
We study in detail the case of
of $d=1$ which describes either the
classical problem of flux lines in a plane with
correlated (columnar) disorder or
interacting fermions (or bosons) in a $d=1$
static random potential. We also give the main features
and the essential physics of the solution for general
$d$. In $d=2$ we study simplified models
relevant to experimental systems
such as the three dimensional vortex lattice with columnar
disorder, disordered quantum bosons in $d=2$,
the pinned Wigner crystal and Charge Density waves.

In a celebrated work \cite{fukuyama_pinning}, Fukuyama and Lee (FL) analyzed
the quantum problem of a one-dimensional CDW with disorder
($d=1$) using a diagrammatic self-consistent method. Physically,
their method, and its further applications \cite{suzumura_scha}
to interacting fermions in $d=1$,
amounts to choose an {\it unknown} classical solution
($\hbar=0$) of the equations of motion in the random potential and
to treat the quantum fluctuations as an harmonic oscillator around this
solution. Since it corresponds to an expansion around an unknown classical
background, the FL method contains an ad-hoc phenomenological parameter.
Although the static correlations are nearly independent of this ad-hoc
parameter, completely different physical behaviors can be obtained
for frequency-dependent quantities depending on its precise value.
To obtain the presumed correct physics the parameter had therefore
to be adjusted by hand. Remarkably, the physics of the
GVM solution parallels the FL theory and allows to establish it on a
firmer footing. As in FL, the physics of the GVM solution
involves an elastic expansion around the classical
solution, but this comes out naturally as the solution of the
variational equations. In addition the GVM allows for a detailed
description of the classical solution. In particular, the
phenomenological parameter
adjusted by hand in FL exactly corresponds to the breakpoint
parameter $u_c$ in the RSB solution ! It is thus
determined by the theory itself.
The GVM therefore provides a more general theoretical framework
which can easily be extended to higher dimensions and more complicated
systems.

In the models considered here, weak pinning is assumed
(disorder is weak and gaussian).
We have not attempted to treat the strong pinning case,
i.e a Poisson distribution of strong pins, although this extension
of the method is possible in principle. For flux line lattices,
disorder can be considered as
gaussian and weak when $R_c > D$ where $D$ is the distance
between columnar defects and $R_c$ the Larkin-Ovchinnikov pinning
length \cite{giamarchi_vortex_short,giamarchi_vortex_long}.
Since columnar defects produced by irradiation are quite strong
pinning centers this corresponds to high fields or temperature. Furthermore
our theory corresponds to a large pinning length $\xi \gg a$, i.e
the elastic limit is assumed, as well as the absence of dislocations.

The plan of the paper is as follows:
in Section~\ref{section2} we introduce the two
classes of models studied
in this paper, quantum problems and
classical problems with correlated disorder.
In Section~\ref{section3} we apply the variational method
to these models and derive the corresponding saddle point equations.
We discuss general features of the solution and
show that, within the GVM, only the zero Matsubara frequency
mode can exhibit replica symmetry breaking.
In Section~\ref{section4} we give the
solution for
a classical model in $d=1$ and $d=2$ describing respectively lines
and vortices in a correlated random potential, as well
a the solution for general $d$.
In Section~\ref{section5} we discuss
the properties of the solution specific to quantum problems, such as
the analytical continuation to real time and we perform the calculation
of the conductivity.
The GVM is applied to the problem
of interacting quantum particles in d=1+1. Conclusions can be found in
section~\ref{section6}.

\section{Models}
\label{section2}

This section introduces the various classical and
quantum models which we study using the variational method.

\subsection{general model}

A general model for an elastic manifold of
internal dimension $d+1$ in
a correlated random potential is:
\begin{equation} \label{general}
 H = \int {d^dx d\tau} {c \over 2}
[ ({\nabla_x }\phi)^2 +  ({\partial_\tau }\phi)^2 ]
+ \int {d^dx d\tau} W((\phi(x,\tau),x)
\end{equation}
$\phi$ can in general be a $N$-component field (to describe
a manifold embedded in $N$ dimensions). The partition
function is $Z=\int  D\phi (x,\tau)  \exp ( - \frac{H}{T} )$.
The disorder potential is Gaussian, correlated in the
direction $\tau$ and the correlator is:
\begin{equation} \label{correlator}
\overline{W(\phi,x) W((\phi',x')} = - C(\phi - \phi') \delta^d(x-x')
\end{equation}
This model describes a large number of classical models and we give
important physical realizations in the next sections.
It also describes the equilibrium statistical mechanics of
quantum problems with disorder. The Hamiltonian
(\ref{general}) can also be viewed as the action of a quantum problem
and $T$ corresponds to $\hbar$.
$\tau=i t$ is the imaginary time of the quantum
problem in the Matsubara representation. $\beta$ is the
inverse temperature of the quantum problem, and
by imposing periodic boundary conditions $\phi(x,\tau=L)=\phi(x,\tau=0)$
one describes bosonic systems, with $L = \beta \hbar$.
Disorder is then time independent which is the natural
situation for a quantum problem. $d=0$ corresponds
to a single quantum particle in a random potential.
Applications to quantum problems are considered in more
details in Section ~\ref{section5}. Note that in this paper
$\beta$ and $T$ are two unrelated parameters.

In order to average over disorder one uses the
replica trick by introducing $k$ replicas of the system $\phi_a$,
$a=1,..k$ and taking the limit $k \to 0$ at the end.
The replicated Hamiltonian averaged over disorder reads:
\begin{equation} \label{general2}
{H_{\text{eff}} \over T}= \int {d^dx d\tau} {c \over 2}
\sum_{a} [ ({\nabla_x }\phi_a)^2 +  ({\partial_z }\phi_a)^2 ]
+ {1\over {2 T}} \int {d^dx d\tau d\tau'}\sum_{a,b}
C(\phi_{a}(x,\tau)-\phi_{b}(x,\tau'))
\end{equation}

\subsection{vortex lattice: all harmonics}

An important physical realization
of the system (\ref{general}) is
the Abrikosov vortex lattice in the presence of columnar
disorder. The columns are
oriented parallel to the $\tau$ direction, the magnetic field being
aligned with the columns. The $d=1$ version of this
problem describes flux lines in a plane pinned by columns of defects
while the $d=2$ version describes the flux lattice with columns
aligned with the direction of the flux lines.

We denote
by $R_i$ the equilibrium position of the vortex
lines labeled by an integer $i$, which defines a perfect lattice of spacing
$a$,
and by $u(R_i,\tau)$ their displacements transverse to the
field.
As will be seen later, the disorder defines a pinning
length $\xi$ (in the transverse direction) such that the
relative displacement of vortices separated by $\xi$
(for equal $\tau$) is of order $a$.
For weak disorder $a/\xi \ll 1$
it is legitimate to assume that $u(R_i,\tau)$ is slowly
varying on the scale of the lattice and to use a continuum elastic energy,
in terms of the continuous variable $u(x,\tau)$.
Impurity disorder is modeled by
a $\tau$ independent
gaussian random potential $V(x)$ with correlations:
$\overline{V(x)V(x')}=W\delta(x-x')$.
The total energy is:
\widetext
\begin{equation} \label{total}
H = \frac12 \int d^dx d\tau \; c[(\partial_\tau u)^2 + (\nabla_x u)^2]
+ \int dx d\tau V(x) \rho(x,\tau)
\end{equation}
\narrowtext
where
the density is $\rho(x,\tau) = \sum_i \delta(x - R_i -u(R_i,\tau))$.
For simplicity we use an isotropic elastic hamiltonian.
Instead of using the displacements $u$, it is more useful to
introduce a smooth ``labelling'' field
\cite{giamarchi_vortex_short,giamarchi_vortex_long}
$\eta(x,\tau) = x - u(\eta(x,\tau),\tau)$.
The density can be rewritten as
\begin{equation} \label{decompos}
\rho(x,\tau) =  \rho_0  det[\partial_\alpha \eta_\beta] \sum_K
e^{i K\eta(x,\tau)}
\approx \rho_0( 1 - \partial_\alpha u_\alpha(\eta(x,\tau),\tau) +
\sum_{K \ne 0} e^{i K x} \rho_K(x))
\end{equation}
where $\rho_K(x)=
e^{-i K \cdot {u}(\eta(x,\tau),\tau)}$ is the usual translational
order parameter defined in terms of the reciprocal lattice vectors
$K$ and $\rho_0$ is the average vortex density.

Using the replica trick on (\ref{total}) the disorder term gives
\begin{equation}
 - \frac{W}{2T} \sum_{a,b} \int dx d\tau d\tau'
\rho^a(x,\tau) \rho^b(x,\tau')
\end{equation}
The above decomposition for the density leads to:
\widetext%
\begin{eqnarray} \label{cardy}
H_{\text{eff}} & = &
\frac12 \int d^dx d\tau \;
\sum_a c[(\partial_\tau u^a)^2 + (\nabla_x u^a)^2]
 \\ & &
- \int d^dx d\tau d\tau'
 \sum_{a,b} [\frac{\rho_0^2 W}{2T}\partial_x
 u^a(x,\tau)\partial_x u^b(x,\tau')
+ \sum_{K\ne 0} \frac{\rho_0^2 W}{2T}
\cos(K (u^{a}(x,\tau)-u^{b}(x,\tau')))]
\nonumber
\end{eqnarray}
\narrowtext%
The range of validity of (\ref{cardy}) is as discussed in
Ref.~\onlinecite{giamarchi_vortex_short,giamarchi_vortex_long}: (i)
elastic limit $a/\xi \ll 1$ (ii) neglect of dislocations,
(which is fully justified in $d=1$). We redefine
$\rho_0^2 W \to W$ in the following.

The slowly varying part of the disorder, i.e the
$q \sim 0$ Fourier components of $V(q)$, couples to the
long wavelength part of the density fluctuations and
gives rise in (\ref{cardy}) to the
quadratic part. The higher Fourier components
of the random potential, specifically those of
Fourier components near $q \sim K$, lead to the cosine
terms. The $q \sim 0$ part of the disorder can easily be
treated, either in the replica form of (\ref{cardy}), or before
averaging by simply redefining:
\begin{equation} \label{redef}
u_\alpha(x,\tau) \to u_\alpha(x,\tau) +  f_\alpha(x,\tau)  ~~~~~
f_\alpha(q,\omega)= \delta_{\omega,0}
\frac{ i \rho_0 q_\alpha V_{q \sim 0} }{c q^2}
\end{equation}
where $V_{q\sim 0}$ is a truncation of the random potential $V(q)$
keeping only the Fourier components close to zero.
Since this redefinition of the $u$
field is $\tau$-independent, it does not affect the cosine terms in
(\ref{cardy}). The $q\sim 0$ part of the disorder decouples
completely from the Hamiltonian (\ref{cardy})
and we thus drop it in the remainder of this paper. It
leads to a simple additional contribution to the correlation functions
which can be computed using (\ref{redef}).

The model (\ref{cardy}) which describes vortex lattices
in correlated disorder is thus a particular case of the
general model (\ref{general},\ref{correlator}).

\subsection{single cosine model}

A further simplification can be obtained by keeping
only the lowest harmonic $K_0$ in (\ref{cardy}). This leads to
another model, which for simplicity we define for a scalar field $\phi$.
This model is studied in detail in this paper:
\begin{equation} \label{cardyos}
{H_{\text{eff}} \over T}= \int {d^dx d\tau} {c \over 2}
\sum_{a} ({\nabla_x }\phi_a)^2 +  ({\partial_z }\phi_a)^2
- \int {d^dx d\tau d\tau'}\sum_{a,b} {W \over {2 T}}
\cos(2(\phi_{a}(x,\tau)-\phi_{b}(x,\tau')))
\end{equation}
It applies directly to CDW and interacting fermions in
$d=1$ as discussed in Section~\ref{section5}. It is
also relevant for flux lines and
correspondence with model (\ref{cardy}) can be made
by defining $u = a \phi/\pi$. As we will see,
in the limit of weak disorder $a/\xi \ll 1$,
the contribution of higher harmonics becomes
irrelevant at large scales, for separations transverse to the magnetic
field.
As for point disorder,
an intermediate ``random manifold'' regime exists at
intermediate scale $x < \xi$ (see Ref. ~\onlinecite{giamarchi_vortex_long}).
For separations along $\tau$, higher harmonics can be included,
as in (\ref{cardy}), but do not change the main physics.

\subsection{domain of application of the models}

In this paper we consider Gaussian disorder,
i.e a large number of very weak pins, rather than
the Poissonian disorder which corresponds to few strong pins.
In addition, we are in the elastic limit of small relative
displacements between nearest neighbors.
Finally in $d=2$, dislocations have been neglected.

Because of the elastic forces (interactions) the
modelization by Gaussian disorder is often appropriate.
Its domain of validity can be estimated at low $T$ as follows.
Let us consider defects of strength $U_0$ separated
by a distance $D$, i.e a density $n \sim 1/D^d$.
The disorder potential is $V(x) = U_0 \sum_i \delta(x-x_i)$
where $x_i$ are the positions of the defects.
Using (\ref{decompos}), and keeping the lowest
harmonic, the coupling to density is
$\sim \rho_0 \sum_i \cos(K_0.(u- x_i))$.
If density of defects is very small, or $U_0$ very strong,
the phase of the lattice adjusts such that $K_0.u = K_0. x_i$
at each defect. In fact one has to balance
the cost in elastic energy with the gain in
potential energy. Assuming that $K_0.u$ varies
by $2 \pi$ over distances of order $\xi$,
the elastic (kinetic) energy is
$ c~\xi^{d-2} $ and the potential energy is,
assuming effectively a Gaussian disorder,
$ U_0 \sqrt{n \xi^d}$. Optimizing over $\xi$,
one finds $\xi \sim (1/n {U_0}^2)^{1/(4-d)}$.
The Gaussian-like regime fails when the
distance $\xi$ becomes {\it smaller} than
the distance $D$ between defects. One then
enters the Poissonian regime where
$\xi$ then saturates at $\xi=D$. From the
expression $\xi/D = (n^{2(2-d)/d}/{U_0}^2)^{1/(4-d)}$
one finds the domain of validity of Gaussian
disorder in any dimensions. In
$d=1$ one has $\xi/D = (n/U_0)^{2/3}$ and the Gaussian
regime holds if the density of defects is not too small
$n \gg U_0$. In $d=2$ one has $\xi/D = 1/U_0$ and
the Gaussian regime holds at any density of defects
provided disorder is small enough $U_0 \ll 1$.
In $d=3$ one has $\xi/D = 1/(n^{2/3} {U_0}^2)$ and the
Gaussian regime holds
for all density and disorder provided they are small.

For models described by a single harmonic,
such as charge density waves, the above argument
always holds. For flux lattices, disorder
can vary at scales $\xi_0$ much smaller than $a$
and higher harmonics
may be important \cite{giamarchi_vortex_long}.
There is an additional length scale, the Larkin
Ovchinnikov length $R_c$, for which relative displacements
are of order $\xi_0$. At weak disorder
$R_c$ and $\xi$ (also called $R_a$) are very different
\cite{giamarchi_vortex_comment}.
A sufficient criterion for Gaussian disorder
to hold is then that $R_c \gg D$.

%For columnar disorder
%$W = U_0^2 b_0^4/d^2$

\section{Variational Method : general properties}
\label{section3}

\subsection{saddle point equations}

We study the statistical mechanics described
by the partition function:
\begin{equation}\label{statmech}
Z=\int  D\phi (x,\tau)  \exp ( - \frac{H}{T} )
\end{equation}
for a system of {\it finite} {\it size} $\beta$ in the $\tau$ direction.
$H$ is the hamiltonian (\ref{general}).
In view of further applications to quantum problems
we impose {\it periodic} boundary conditions in the $\tau$ direction. For
infinitely thick vortex systems of length $L=\beta \rightarrow
\infty$ boundary conditions should not matter. However,
imposing definite boundary conditions
and working first on a finite size sample is crucial for
obtaining well-defined equations in the case
of correlated disorder.

Introducing replicas and averaging over disorder the model
becomes (\ref{general2}). We now use the variational method
as in Ref.~\onlinecite{mezard_variational_replica}. This is done by choosing
a Gaussian variational Hamiltonian:
\begin{equation} \label{variat}
H_0 = {1 \over 2} \int {d^dq \over (2 \pi)^d}
\sum_{n} G^{-1}_{ab}(q,\omega_n)
\phi_a(q,\omega_n) \phi_b(-q,-\omega_n)
\end{equation}
where the discrete summation is over $\omega_n=2 \pi n/\beta$
analogous to Matsubara frequencies. Note that
$\frac{1}{\Omega} \sum_{q} \to \int \frac{d^dq}{(2 \pi)^d} $ and
$\frac{1}{\beta} \sum_{\omega_n} \to \int \frac{d\omega}{2 \pi} $
in the large volume ($\Omega$, resp. $\beta$) limit.

By the same arguments as in Ref.~\onlinecite{mezard_variational_replica}
this method is exact for a choice of correlator
of the form $C(\phi - \phi')= N V((\phi - \phi')^2/N)$
in the limit $N \to \infty$. This corresponds to a manifold
in infinite dimensions, either a classical manifold awith correlated disorder
or a quantum manifold with static disorder.
We obtain the variational free energy $F_{\text{var}}=F_0+\langle H-H_0
\rangle_{H_0}$ as
\begin{eqnarray} \label{enervar}
F_{\rm var} & = &
\frac{T}2 \sum_a \sum_{q,\omega_n} \left( c (\omega^2_n + q^2)
G(q,\omega_n)_{aa} - [\ln(T G)]_{aa}  \right) \nonumber \\
 & & + \frac1{2T} \sum_{ab} \int d^dx \int_0^{\beta} \int_0^{\beta}
d\tau d\tau' ~V(B_{ab}(x=0,\tau-\tau'))
\end{eqnarray}
where
\begin{eqnarray}
B_{ab}(x,\tau) & = &  \langle [\phi_a(x,\tau) - \phi_b(0,0)]^2 \rangle  \\
            & = &  \frac{T}\beta \int {d^dq \over {(2\pi)}^d} \sum_n
(G_{aa}(q,\omega_n) + G_{bb}(q,\omega_n) - 2 \cos(qx+\omega_n \tau)
G_{ab}(q,\omega_n))
\nonumber
\end{eqnarray}
In general the function $V(B)$ which appears in the variational energy is
defined as:
\begin{equation}
\langle C(\phi) \rangle = V(\langle \phi^2 \rangle)
\end{equation}
where $\langle ..\rangle$ denotes an average over any (one-component)
gaussian random variable $\phi$. For the case of model
(\ref{cardyos}) $V(B)= - W \exp(-2 B)$ and
$V(B) = -  W \sum_K \exp{- K^2 B/2 }$ for model
(\ref{cardy}). Defining the self-energy $G^{-1}_{ab}=  c q^2 \delta_{ab} -
\sigma_{ab}$, and $G_c^{-1}(q) = \sum_b G_{ab}^{-1}(q)$
we obtain by minimization
of the variational free energy $F_{\text{var}}$ the saddle point equations:
\begin{eqnarray} \label{bebe}
G_c^{-1} & =& c (q^2 + \omega_n^2) + \frac{2}T \int_0^\beta d\tau
               (1-\cos(\omega_n \tau)) [ V'(B_{aa}(0,\tau))
                + \sum_{b\ne a} V'(B_{ab}(0,\tau))] \nonumber \\
\sigma_{a\ne b} & = & \frac{2}T \int_0^\beta d\tau \cos(\omega_n\tau)
                   V'(B_{ab}(0,\tau))
\end{eqnarray}
Taking the limit of the number of replicas $k \to 0$
and considering the most general solution, i.e.
a replica symmetry broken solution, we denote
$\tilde{G}(q,\omega_n)=G_{aa}(q,\omega_n)$
and parameterize $G_{ab}(q,\omega_n)$
by $G(q,\omega_n,u)$ where $0<u<1$, and similarly for $B_{ab}(x,\tau)$
and $B(x,\tau,u)$. Equ.~(\ref{bebe}) and
the algebraic rules for hierarchical matrices
\cite{mezard_variational_replica} give:
\begin{eqnarray} \label{eqbase}
G_c^{-1} & =& c (q^2 + \omega_n^2) + \frac{2}T \int_0^\beta d\tau
               (1-\cos(\omega_n \tau)) [ V'(\tilde{B}(\tau))
                -\int_0^1 du V'( B(\tau,u))] \nonumber \\
\sigma(\omega_n,u) & = & \frac{2}T \int_0^\beta d\tau
                         \cos(\omega_n\tau) V'(B(\tau,u))
\end{eqnarray}
with
\begin{eqnarray}
B(\tau,u) & =& \frac{2T}{\beta}\sum_n \int \frac{d^d q}{(2\pi)^d}
[\tilde{G}(q,\omega_n) - G(q,\omega_n,u) \cos(\omega_n \tau)] \nonumber
\\
\tilde{B}(\tau) & = & \frac{2T}{\beta}\sum_n \int_0^\beta
\frac{d^d q}{(2\pi)^d}
\tilde{G}(q,\omega_n) (1 - \cos(\omega_n \tau))
\end{eqnarray}
where we denote $\tilde{B}(\tau)=\tilde{B}(x=0,\tau)$ and
$B(\tau,u)=B(x=0,\tau,u)$.
A replica symmetric solution would correspond quantities constant in $u$
in (\ref{eqbase}), and is analyzed in Appendix~\ref{appendixA}.
Note that the connected part $G_c^{-1}(q,\omega)$, is renormalized
by correlated disorder, while it is not the case for point disorder
\cite{giamarchi_vortex_long}.
This is a crucial difference which leads to localization and
stiffening of the tilt modulus.

\subsection{general features of the solution}

A general and important property of quantum problems with disorder
is that off-diagonal quantities like $B_{a\ne b}(x,\tau)$
are in fact always independent of $\tau$.
The general argument \cite{subir_replicas} is that
in each realization of the random potential W, the disorder does not depend
on $\tau$. Therefore before averaging over disorder:

\begin{equation} \label{timeind}
G_{ab,W} = \langle \phi_a(x,\tau) \phi_b(0,0)\rangle =
           \langle \phi_a(x,\tau)\rangle \langle\phi_b(0,0)\rangle =
           \langle \phi_a(x,0)\rangle \langle \phi_b(0,0)\rangle
\end{equation}
It is important to note that such a property crucially depends on
the assumption that the hamiltonian is $\tau$-independent and
on the fact that equilibrium has being attained. This is the case
considered here. Perturbations which explicitly depend on imaginary or real
time destroy this property and may lead to more complex
behavior, as discussed in Section~\ref{section6}.

It is easy to check on the saddle point equations (\ref{eqbase})
that they indeed admit a solution such that off-diagonal quantities
are $\tau$-independent. For this solution the saddle point equations
simplify greatly:

\begin{eqnarray} \label{eqbasesimple}
G_c^{-1} & =& c (q^2 + \omega_n^2) + \frac{2}T \int_0^\beta d\tau
               (1-\cos(\omega_n \tau)) [ V'(\tilde{B}(\tau))
                -\int_0^1 du V'( B(u))] \nonumber \\
\sigma(\omega_n,u) & = & \frac{2 \beta}T V'(B(u)) \delta_{n,0}
\end{eqnarray}
with
\begin{eqnarray} \label{form}
B(u) & =&
\frac{2 T}\beta \sum_{n\ne 0} \int \frac{d^dq}{(2\pi)^d} G_c(q,\omega_n)
+
\frac{2T}{\beta} \int \frac{d^d q}{(2\pi)^d}
[\tilde{G}(q,\omega_n=0) - G(q,\omega_n=0,u)] \nonumber \\
\tilde{B}(\tau) & = & \frac{2T}{\beta}\sum_n \int_0^\beta
\frac{d^d q}{(2\pi)^d} G_c(q,\omega_n) (1 - \cos(\omega_n \tau))
\end{eqnarray}
where we denote $B(\tau,u)=B(u)$. The simplifications are the following.
Since $B(\tau,u)$ is independent of $\tau$ the self energy
$\sigma(\omega_n,u)$ vanishes for all modes excepted $\omega_n=0$.
One also has $G(q,\omega_n,u)=0$ for $\omega_n \neq 0$, and thus
$\tilde{G}(q,\omega_n)= G_c(q,\omega_n)$ for $\omega_n \neq 0$.

An obvious property of the above saddle point is thus
that replica symmetry breaking is confined to the
mode $\omega_n=0$. The equation for this mode:
\begin{equation} \label{short}
\sigma(\omega_n,u) = \delta_{n,0} \sigma(u) = \frac{2 \beta}{T} \delta_{n,0}
V'(B(u))
\end{equation}
is {\it identical} to the one for a model with
pointlike disorder in $d$ dimensions, studied in
\cite{mezard_variational_replica,giamarchi_vortex_short,giamarchi_vortex_long}.
The only (important) difference is that the potential is effectively
multiplied by $\beta$, i.e the total length along the $\tau$ direction.
This is why a single mode ($\omega_n=0$) can lead to finite contributions
to correlation functions in the limit $\beta \to \infty$. In that limit
the corresponding model
with point disorder in $d$ dimensions can, in some sense, be considered as
effectively at zero temperature $\sim T/\beta$.

Known results
\cite{mezard_variational_replica,giamarchi_vortex_short,giamarchi_vortex_long}
can thus be transposed to study the zero mode $\omega_n=0$.
For the potentials with power law correlators \cite{mezard_variational_replica}
$V(x)=g  x^{1-\gamma}/(2(1-\gamma))$
there are two generic cases: (i) long range potentials,
if $\gamma (1 - d/2) < 1$, for which one has full replica symmetry breaking
and (ii) short range potentials, if $\gamma (1 - d/2) > 1$,
for which one has one step replica symmetry breaking
(and a transition to a high temperature Replica Symmetric (RS) phase).
In the case of the single cosine model (\ref{cardyos})
there is
\cite{giamarchi_vortex_short,giamarchi_vortex_long,korshunov_variational_short}
a one step solution
for $d \leq 2$ and a full RSB solution for $d > 2$.
Potentials $V(x)$ which are realistic
to describe media with several scales, such as the vortex lattice
(\ref{cardy}), will include several of the above regimes
%% FOLLOWING LINE CANNOT BE BROKEN BEFORE 80 CHAR
\cite{giamarchi_vortex_short,giamarchi_vortex_long,bouchaud_variational_vortex}.

The general way to study (\ref{short}) is as follows. There is
a breakpoint $u_c$ such as $\sigma(u)$ is constant for $u>u_c$.
For $u>u_c$ the inversion rules applied to (\ref{form}) give:
\begin{equation} \label{inversion0}
B(u > u_c)=B(u_c)=\frac{2 T}\beta \int \frac{d^dq}{(2\pi)^d} \sum_{n\ne 0}
        (G_c(q,\omega_n) + \frac{2 T}\beta \int \frac{d^dq}{(2\pi)^d}
        \frac1{q^2+\Sigma_1}
\end{equation}
where by definition $\Sigma_1=[\sigma](u_c)$ and
the function $[\sigma](u)=u\sigma(u)-\int_{0}^{u} dv \sigma(v)$.
One searches for a full RSB solution by applying to (\ref{form})
the inversion formula for $u < u_c$:
\begin{equation} \label{inversion}
B(u) = B(u_c) + \frac{2 T}\beta \int_{u}^{u_c} dv \int
\frac{d^dq}{(2 \pi)^d}
{\sigma' (v) \over {G_c(q,\omega_n=0)^{-1} +
[\sigma](v))}^2 }
\end{equation}
Differentiating (\ref{short}) with
respect to $u$, using (\ref{inversion}) and
$[\sigma]'(u)= u \sigma'(u)$ one can then divide
by $\sigma'(u) \neq 0$ since full RSB is assumed. The
quantities $B(u)$, $[\sigma](u)$ and $\sigma(u)$ are then
determined by elimination from the resulting system:
\begin{eqnarray} \label{thesame}
1 & = & - 4 V''(B(u)) \int \frac{d^d q}{(2\pi)^d}
             \frac{1}{(c q^2 + [\sigma](u))^2} \nonumber \\
\sigma(u) & = &  \frac{2 \beta}{T} V'(B(u))
\end{eqnarray}
and the definition of $[\sigma](u)$.
These equations, which govern the mode $\omega_n=0$,
are {\it identical}
to the one obtained previously for pointlike disorder in $d$ total dimension,
up to a rescaling of $u$ by $\beta$,
i.e $u$ is smaller by a factor of $\beta$ in this problem.
$[\sigma]$ and $B$ do not scale but $\sigma$ is larger by a factor $\beta$.
The full RSB solution for the power law potential is thus
$[\sigma](u)=  \Sigma_1 (\beta u/u_c)^{2 / \theta}$ with
$\theta = (2 + (d-2) \gamma)/(1+\gamma)$.
Once the equation is solved for $u < u_c$,
and provided that a full RSB solution exists,
which is the case when $\theta >0$,
the unknown
constants $\Sigma_1$ and $B(u_c)$ are determined by
matching or equivalently
by elimination in:
\begin{eqnarray} \label{thesame3}
1 & = & - 4 V''(B(u_c)) \int \frac{d^d q}{(2\pi)^d}
             \frac{1}{(c q^2 + \Sigma_1)^2} \nonumber \\
B(u_c)& = & \frac{2 T}\beta \int \frac{d^dq}{(2\pi)^d} \sum_{n\ne 0}
        G_c(q,\omega_n) + \frac{2 T}\beta \int \frac{d^dq}{(2\pi)^d}
        \frac1{q^2+\Sigma_1}
\end{eqnarray}

For the single cosine model (\ref{cardyos}) which we focus
on here, the above system (\ref{thesame}) simplifies into:
\begin{equation} \label{thesame2}
1  = \beta \sigma(u) \int \frac{d^d q}{(2\pi)^d}
             \frac{1}{(c q^2 + [\sigma](u))^2}
\end{equation}
and the solution of (\ref{thesame}) is
\cite{giamarchi_vortex_short,giamarchi_vortex_long}
in $d > 2$ $[\sigma](u)=  (\beta u/u_0)^{2 / \theta}$ and
$u_0 = 8 T c_d c^{-d/2}/(4-d)$. The exponent
$\theta=d-2$ characterize the scaling of
the fluctuations of the energy of the metastable
states with distance, $\Delta F \sim x^\theta$.
A pair of states characterized by $u$ in Parisi's
hierarchy are typically $x$ apart in space and
$\Delta F= T/u$ apart in energy. The large scale behavior
is thus controlled by small $u$.

In $d \leq 2$, the single
cosine model does not admit a full RSB solution.
Instead one finds a one step RSB solution
with $\sigma(u) = 0$ for $u < u_c$ and
$\sigma(u) = \sigma(u_c)$ for $u \geq u_c$. One then has:
\begin{eqnarray}
[\sigma](u) & = & 0 \qquad u < u_c  \label{onestep} \\
{[\sigma]}(u) & = & \Sigma_1 = \beta u_c \frac{2}T V'(B) \qquad u > u_c
\label{onestep5}
\end{eqnarray}
where we have used (\ref{short}) and the fact that $\Sigma_1=[\sigma](u_c) =
u_c
\sigma(u_c)$ for such a solution. One has also $B(u)=\infty$ for $u < u_c$
and we denote $B(u > u_c)=B$.
For the single cosine model $\Sigma_1 = \beta u_c \frac{4W}T e^{-2 B}$.
To determine $\Sigma_1$, $B$ and $u_c$ in the one step case,
one can use only equations (\ref{onestep5},\ref{inversion}). One equation is
still missing. Before we discuss the additional equation needed
to determine $u_c$, let us further simplify the saddle point equations.

Irrespective of whether the solution
is one step or full RSB, one can always rewrite
the variational equations as:
\begin{eqnarray} \label{fulleq1}
G_c^{-1}(q,\omega_n) & = &  c(q^2 + \omega_n^2)  +
\Sigma_1(1-\delta_{n,0}) + I(\omega_n) \\
I(\omega_n) & = & \frac{2}T \int_0^\beta d\tau
           (1-\cos(\omega_n \tau)) (V'(\tilde{B}(\tau)) - V'(B))
\label{iom0}
\end{eqnarray}
with
\begin{eqnarray} \label{fulleq2}
B & = & \frac{2 T}\beta \int \frac{d^dq}{(2\pi)^d} \sum_{n\ne 0}
        G_c(q,\omega_n) + \frac{2 T}\beta \int \frac{d^dq}{(2\pi)^d}
        \frac1{q^2+\Sigma_1} \\
\tilde{B}(\tau) & = & \frac{2T}{\beta}\sum_n
\frac{d^d q}{(2\pi)^d}
G_c(q,\omega_n) (1 - \cos(\omega_n \tau)) \nonumber
\end{eqnarray}
where $B(u > u_c)=B$.
To obtain (\ref{fulleq1}) from (\ref{eqbase}) we have splitted:
\begin{equation}
\int_0^1 du V'(B(u)) = (1-u_c) V'(B) + \int_0^{u_c} du V'(B(u))
= V'(B) - \frac{T}{2 \beta} \Sigma_1
\end{equation}
and used the equation (\ref{short}) and the definition
of $[\sigma](u)$. The
form (\ref{fulleq1}) is convenient because
when $\beta$ becomes large,
$\tilde{B}(\tau)$ converges at large $\tau$
towards $B$ and thus $I(\omega_n)$ in (\ref{iom0})
goes to zero at small $\omega_n$. One has:
\begin{equation}
B - \tilde{B}(\tau) =
\frac{2 T}\beta \sum_{n} \int \frac{d^dq}{(2\pi)^d}
\cos(\omega_n \tau)
\frac1{c q^2+ c \omega_n^2 + \Sigma_1 + I(\omega_n)}
\end{equation}
and thus $\lim_{\tau \to \infty} \lim_{\beta \to \infty} \tilde{B}(\tau) = B$.

One thus sees on (\ref{fulleq1}) that disorder produces a ``mass'' term
$\Sigma_1$
and thus {\it localization} of the elastic manifold. The mechanism by
which it is generated is subtle, however. Such term
cannot exist at $\omega_n=0$ as can be seen from (\ref{eqbase}).
Indeed
$G_c^{-1}(q=0, \omega=0)=0$ as it must from
translational invariance after averaging over disorder.
As soon as $\omega_n = 2 \pi n/\beta \neq 0$ the first
term in (\ref{fulleq1}) becomes non zero equal to $\Sigma_1$
while the second one
is a smooth function of $\omega_n$ and therefore remains
small near $\omega=0$. Thus we find, and observe in a
numerical solution of equations (\ref{fulleq1},\ref{fulleq2}),
that in the limit $\beta \to \infty$, $G_c^{-1}(q, \omega)$
develops a discontinuity at $\omega=0$. In fact
the only way to
obtain a {\it finite} localization length in the limit
$\beta \to \infty$ is to break replica symmetry.
The formula
(\ref{fulleq2}) for $B$ shows that $\Sigma_1$ plays the role
of an effective mass which makes $B$ finite
in $d \leq 2$. There is no such mechanism in the
replica symmetric solution, which cannot describe properly
the localized phase and is studied in
detail in Appendix~\ref{appendixA}.

Remarkably, Eqs.
(\ref{fulleq1}-\ref{fulleq2}) now forms a closed system
of equations, depending only on one additional
quantity $\Sigma_1$, and
do not depend on how the replica symmetry is
broken or on the detailed structure of the solution
$B(u)$ for $u<u_c$. We will discuss further this property
in Section~\ref{section6}.

\section{Elastic manifolds in correlated potentials}
\label{section4}

It is possible to solve analytically the saddle point equations
(\ref{fulleq1}-\ref{fulleq2}) in the limit of low temperature
$T \to 0$. Such solution, as we will show, contains the essential
physics of the localized phase. It is important to retain the finite
temperature only for few quantities that we will discuss explicitly.

In the limit when $T\to 0$,
one can expand the potential in the expression (\ref{iom0})
of $I(\omega_n)$. One gets a self-consistent equation
for $I(\omega_n)$:
\begin{equation} \label{iom}
I(\omega_n) = - 4 V''(0) \int \frac{d^dq}{(2\pi)^d} \left(
\frac{1}{c q^2 + \Sigma_1} -
\frac{1}{c q^2 + \Sigma_1 + c \omega_n^2 + I(\omega_n)} \right)
\end{equation}
It is important to note that the low-T expansion of the
potential can be performed safely only
on the form (\ref{fulleq1}) of the saddle point
equations, $\Sigma_1$ being fixed. It would be incorrect
to expand directly (\ref{eqbasesimple}) since $B(u)$ diverges
at small $u$ and the low-T expansion breaks down.
We now study different cases.

\subsection{Solution for the cosine model in $d=2$} \label{degal2}

In $d=2$ the solution for the single cosine
model (\ref{cardyos}) is one step but marginally so. The
value of $u_c$ is determined unambiguously either from
the full RSB solution in the limit
$d \to 2^{+}$, or by minimizing the free energy $F_{\rm
var}$ in (\ref{enervar}). This is done similarly to the case of point
like disorder \cite{giamarchi_vortex_long}. For model
(\ref{cardy}) in the continuous limit:
\begin{equation} \label{betauc}
\beta u_c  =  \frac{T K_0^2}{4 \pi c}
\end{equation}
and thus $\beta u_c = T/\pi c$ for model  (\ref{cardyos}).
Note that there is a transition to a high temperature
phase where disorder is irrelevant at $T=T_c = \pi \beta c$.
For $\beta$ not very small this temperature however is
very high and in realistic systems other mechanisms,
e.g dislocation induced melting,
will cause a transition out of the glass phase before
this one occurs.
Thus we confine our study to the low temperature glass phase
$T \ll T_c$.

One then obtains from (\ref{iom})
\begin{equation}
I(\omega_n) = 4\pi c \Sigma_1 \int \frac{d^2q}{(2\pi)^2}
[\frac1{c q^2+\Sigma_1}
              - \frac1{c q^2 + \Sigma_1 + c \omega_n^2 + I(\omega_n)}]
\end{equation}
which leads to, in the limit where the cutoff goes to infinity
\begin{equation} \label{deto}
I(\omega_n) = \Sigma_1 \log(1 + \frac{I(\omega_n)+ c \omega_n^2}{\Sigma_1})
\end{equation}
The solution of (\ref{deto}) can be written as
\begin{equation} \label{scaling}
I(\omega_n)  =  \Sigma_1 f(\sqrt{c} |\omega_n |/\sqrt{\Sigma_1})
\end{equation}
$\Sigma_1$ defines a natural length scale $\xi = \sqrt{c/\Sigma_1}$.
$\xi$ is called the pinning length and corresponds to the
scale above which the
long range order is destroyed due to disorder.

For small $\omega_n$ the solution of (\ref{deto}) is
\begin{equation}
I(\omega_n) \sim \sqrt{2 c \Sigma_1} | \omega_n |
\end{equation}
and for large $\omega_n$
\begin{equation}
I(\omega_n) \sim \Sigma_1 \log(c \omega_n^2/\Sigma_1)
\end{equation}
This formula is valid up to the cutoff $\omega_n \sim \Lambda$.
At large $\omega$, $I(\omega) \to I(\infty)= \int \frac{d^2q}{(2\pi)^2}
\frac1{c q^2+\Sigma_1}$.
$\Sigma_1$ is determined from (\ref{thesame3},\ref{betauc},\ref{onestep5}).
At low $T$ one finds $\Sigma_1 = 4 W/\pi$.

The general expression for the correlation functions
for a one step solution is:
\begin{eqnarray} \label{funccorstep}
\tilde{B}(x,\tau) & = & \overline{\langle[\phi(x,\tau)-\phi(0,0)]^2\rangle}=
\frac{2 T}\beta \sum_n \int \frac{d^d q}{(2\pi)^d}
         \frac1{c \omega_n^2+c q^2+\Sigma_1 +
         I(\omega_n)}(1-\cos(q x + \omega_n \tau))
\nonumber \\
& & +\frac{2 T}{\beta u_c} \int\frac{d^d q}{(2\pi)^d}
                \frac{\Sigma_1}{c q^2(c q^2+\Sigma_1)}(1-\cos(q x))
\end{eqnarray}
At $x=0$, $\tilde{B}(0,\tau)$ goes to a constant when $\tau\to\infty$.
This describes the localization of a flux line due to the correlated
disorder. One defines a localization length $l_{\perp}$ as:
\begin{equation} \label{loclength}
l_{\perp}^2 = \frac{a^2}{\pi^2}
\lim_{\tau \to \infty}
\overline{\langle[\phi(x,\tau)-\phi(0,0)]^2\rangle}
\end{equation}
remembering that we have defined the displacements as
$u=a \phi/\pi$. One finds
from (\ref{funccorstep}):
\begin{equation} \label{lperp}
l_{\perp}^2 = \frac{a^2}{\pi^2} B = \frac{a^2}{\pi^2} \frac{2 T}{\beta}
\sum_n \int \frac{d^d q}{(2\pi)^d}
\frac1{c \omega_n^2+c q^2+\Sigma_1 + I(\omega_n)}
\end{equation}
In $d=2$ one finds:
\begin{equation}
l_{\perp}^2 = l_{T}^2 - \frac{a^2}{\pi^2} \frac{2 T}{c}
\sqrt{\frac{\Sigma_1}{c}}
\frac1{4 \pi^2}
\int_0^{\infty} dx
\ln (1 + \frac{1 + f[x]}{x^2} )
\end{equation}
where $l_{T}^2=2T a^2/(c \pi^2) \int \frac{d\omega d^2 q}{(2\pi)^3}
(q^2+\omega^2)^{-1}$ is the Lindemman length. This
formula can be used to compute the upward shift in the
melting temperature due to correlated disorder using
a Lindemman criterion.

On the other hand for fixed $\tau$, the $\omega_n=0$ mode
dominates at large $x$ and $\tilde{B}(\tau,x)$ grows as:
\begin{equation} \label{cor2}
\tilde{B}(x,\tau) \sim \frac{T}{c \beta u_c \pi} \log(x/\xi)
= \log(\frac{x}{\xi})
\end{equation}
The prefactor in (\ref{cor2}) is universal and lead to
a power law decay of the translational order parameter
\cite{giamarchi_vortex_long}. The additional
contribution of the $q\sim 0$ Fourier components of the random
potential can be computed from (\ref{redef}) and gives
a small non universal correction.

One can define the $\omega$-dependent
(i.e $q_z$ dependent, in usual notations) tilt modulus $c_{44}(\omega)$:
\begin{equation} \label{tiltmod}
\frac{T}{c_{44}(\omega)} = \lim_{q \to 0}
\langle \partial_\tau \phi \partial_\tau \phi \rangle
= \lim_{q \to 0} \omega^2
\langle \phi(q,\omega) \phi(-q,-\omega) \rangle
\end{equation}
Using equation (\ref{fulleq1}) one finds:
\begin{equation}
\frac1{c_{44}(\omega)} = \frac{\omega^2}{c \omega^2 + \Sigma_1 + I(\omega)}
\end{equation}
a formula which holds in any dimension. Thus we find that
$c_{44}=c_{44}(\omega \to 0)$ diverges in the Bose
glass phase, a signature of the transverse Meissner effect
\cite{nelson_columnar_long}.

We find from the GVM three important properties of the Bose Glass
phase: the finite localization length $l_{\perp}$,
the decay of the translational order in the direction transverse
to the columnar defects, and the divergence of the tilt modulus.

\subsection{Solution for the cosine model in $d=1$} \label{sol1d}

In $d=1$ thermal fluctuations are much stronger and in
the absence of disorder correlation functions grow
logarithmically at large distances:
\begin{equation} \label{sans}
\tilde{B}(x,\tau) \sim  \frac{T}{2 \pi c} \ln[(x^2 + \tau^2)/a^2]
\end{equation}
In the presence of
disorder one expects a low temperature localized phase were
disorder is relevant and a replica symmetric high temperature phase
were thermal fluctuations dominate. The transition at $T_{BG}=
6 \pi c/K_0^2$ (at infinitesimal disorder) can be seen
on the RS solution (see appendix~\ref{appendixA})
and is found to coincide with the results
\cite{giamarchi_loc} of the RG. Eq. (\ref{iom}) leads to:
\begin{equation} \label{fukulee}
I(\omega_n) = - \frac{4 V''(0)}{2 \sqrt{c}}
 [ \Sigma_1^{-1/2} - (\omega_n^2 + \Sigma_1
              + I(\omega_n))^{-1/2} ]
\end{equation}
To solve this equation one needs to know $\Sigma_1$, and thus
determine $u_c$. Most of the physical properties, such as
the relative displacements in the transverse direction,
the localization along the columns, or the vanishing
of the tilt modulus are independent of the precise
value of $u_c$. However, quantities for which
the asymptotic $\tau$-dependence is important are crucially
dependent on $u_c$.

The standard determination of $u_c$ in the GVM
in the one step case is
by minimizing the free energy $F_{\rm
var}$ in (\ref{enervar}) with respect to $u_c$. This can be done
similarly to the case of point like disorder
\cite{giamarchi_vortex_long}.
The result for the cosine model in $d < 2$ is:
\begin{equation}
\beta u_c =  4 T \frac{2-d}{d} j_d \Sigma_1^{(d-2)/2}
\end{equation}
where $j_d  =  \int_{-\infty}^{+\infty} \frac{d^dq}{(2\pi)^d} (q^2+1)^{-1}$.
Thus $\beta u_c = 2 T \Sigma_1^{-1/2} c^{1/2}$ for $d=1$ ($j_1=1/2$).
A similar result for a general short range potential
can be obtained and involves $V'(B)$.
For the cosine model if one uses the value of $u_c$ obtained
from minimization one gets:
\begin{equation} \label{fukuleed1}
I(\omega_n) = \Sigma_1^{3/2} [ \Sigma_1^{-1/2} - (\omega_n^2 + \Sigma_1
              + I(\omega_n))^{-1/2} ]
\end{equation}
Expanding the second term in powers of $I(\omega_n)$ one finds that, unlike
what happens in $d=2$, the linear order does not vanish. As a consequence
the solution of (\ref{fukuleed1}) behaves as $I(\omega_n) \sim
\omega_n^2$ at small $\omega_n$. This solution would lead
to an exponential decay of $\tilde B(\tau)$ with $\tau$ towards
its asymptotic value. For the classical problem this
type of behavior cannot be excluded, though it
would lead to a surprising difference between
$d=2$ and $d=1$. However, for the associated quantum problem there is strong
evidence that this solution gives {\it incorrectly} the physics,
for example it leads to unphysical result for the conductivity
as we discuss in Section~\ref{section5}. We will thus
not explore this solution further.

A look at (\ref{iom}) shows that the general condition for the vanishing
of the linear term in the expansion in $I(\omega_n)$, leading to
$I(\omega_n) \sim  |\omega_n|$ at small $\omega_n$ reads:
\begin{equation} \label{replicon2}
1 = - 4 V''(B) \int \frac{d^dq}{(2\pi)^d} \frac{1}{(c q^2 + \Sigma_1)^2}
\end{equation}
One recognize this condition as the condition of {\it marginal
stability}, i.e the vanishing of the replicon eigenvalue.
As discussed in the next section, such a criterion is true at
all temperatures.
This condition involves the second derivative of the potential
and is thus different from the equation obtained by minimization
of the free energy over $u_c$. The marginal stability condition
is known to arise naturally in the problem of the {\it Langevin dynamics}
of models which have a one step RSB solution in the statics
(see discussion in Section~\ref{section6}).
This is the condition that we choose here, on physical grounds. For $d
= 2$ this condition coincides with the one obtained by minimizing the
free energy studied in section~\ref{degal2}.

With this choice of $u_c$, $\beta u_c = T \Sigma_1^{-1/2} c^{1/2}$
(\ref{fukuleed1}) becomes:
\begin{equation} \label{fukuleed2}
I(\omega_n) = 2 \Sigma_1^{3/2} [ \Sigma_1^{-1/2} - (\omega_n^2 + \Sigma_1
              + I(\omega_n))^{-1/2} ]
\end{equation}
The solution of (\ref{fukuleed2}) can be written as
\begin{equation} \label{scaling2}
I(\omega_n)  =  \Sigma_1 f(\sqrt{c} |\omega_n |/\sqrt{\Sigma_1})
\end{equation}
where
\begin{eqnarray} \label{limits}
f(x) & \sim & \frac{2}{\sqrt3} x - \frac49 x^2 \qquad {\rm when}\; x \sim 0 \\
f(x) & = & 2 - \frac1{\sqrt{x}} \qquad {\rm when}\;  x \to \infty
\end{eqnarray}
Using the above solution and formulae (\ref{funccorstep}) one can compute
the correlation functions.
For fixed $\tau$, the $\omega_n=0$ mode
dominates at large $x$ and $\tilde{B}$ grows as:
\begin{equation} \label{growth}
\tilde{B}(\tau,x) \sim \frac{T}{\beta u_c} \frac{|x|}{c} = \frac{|x|}{\xi}
\end{equation}
where we have used the value $\beta u_c = T/\sqrt{c \Sigma_1}$
and $\xi$ is the pinning length.

In the limit
of small disorder $W$, and away from the transition,
$\Sigma_1$ and thus $\xi$
can be computed as a function of the
temperature $T$ (for $\beta\to\infty$).
Introducing an ultraviolet cutoff $q_{max}=\Lambda$, and rescaling the
variables using (\ref{scaling2}) one gets from (\ref{fulleq2}):
\begin{equation} \label{valb}
B = 2 T \int^{\Lambda/\sqrt{\Sigma_1}}
              \frac{dq d\omega}{(2\pi)^2} \frac1{c q^2 + c \omega^2
             + 1 + f(\sqrt{c} |\omega|)} \sim \frac{T}{\pi c}
               \log( \sqrt{c} \Lambda/\sqrt{\Sigma_1})
\end{equation}
where one has used (\ref{limits}). Inserting the value of $B$ (\ref{valb})
in (\ref{onestep5}) and using $\beta u_c = T/\sqrt{c \Sigma_1}$
one gets:
\begin{equation}\label{pinlength}
c/\xi^2 = \Sigma_1 =
(4 W/\sqrt{c})^{\frac2{3-2T/\pi c}}\Lambda^{-\frac{4T/\pi c}{3-2T/\pi c}}
\end{equation}
This formula for the weak disorder behavior
is similar to the one that was derived for the pinning of
charge density waves by Fukuyama and Lee
\cite{fukuyama_pinning} (for $T=0$) and extended to take into
account the quantum fluctuations (here the thermal fluctuations $T >0$) by
Suzumura and Fukuyama \cite{suzumura_scha}.
The formula
(\ref{valb}) is not identical however to the one of
Suzumura and Fukuyama since it involves an integration
over $I(\omega)$ which it determined self-consistently
in our theory. In Ref.~\onlinecite{suzumura_scha}
the corresponding $I(\omega)$ is zero, which if
taken seriously would lead to unphysical results
for quantities such as $c_{44}$ or $\sigma(\omega)$.
In addition, contrarily to their solution, it does not
contain any unknown prefactor. The scaling with disorder
of (\ref{pinlength}) coincides with the result from the
renormalization group analysis \cite{giamarchi_loc}.
The phase transition at $T=T_{BG}$ appears on these formula.
Additional calculations at finite temperature
have been performed in Appendix~\ref{onedim}.

At $x=0$, $\tilde{B}(0,\tau)$ goes to a constant when $\tau\to\infty$.
This describes the localization of the lines due to the correlated
disorder. One finds the expression of the localization radius:
\begin{equation} \label{locrad}
\l_{\perp}^2 = \frac{a^2}{\pi^2}  \frac{T}{ \pi c}
\ln (\xi/a)
\end{equation}
A comparison with the result (\ref{sans}) shows that
$\tilde{B}(\tau)$ grows until $\tau \sim \xi$ when
it saturates at $l_{\perp}^2$. Note the difference
between the two lengths $l_{\perp}$ and $\xi$.
At the transition one expects that
$\xi$ diverges as $\xi \sim \exp(b/(T_{BG}-T)^\alpha)$ with
$\alpha = 1/2$ from the RG \cite{giamarchi_loc}. Thus
relation (\ref{locrad}) would predict that:
\begin{equation}
\l_{\perp} \sim \frac{1}{(T_{BG}-T)^{\alpha/2}}
\end{equation}
which could be measured in numerical simulations.

For the single cosine model
in $d=1$ and for $\beta$ finite, one does not expect
a real phase transition (at $T>0$) since the system is
quasi-one dimensional. The GVM gives however
a one step RSB solution at finite large enough $\beta$,
$\beta \sim T $. This
solution is of the kind discussed in Appendix~\ref{variational}:
it is discontinuous and disappears when disorder is decreased below a finite
threshold. It is an artifact of the GVM and should not be interpreted
as a genuine phase
transition in $d=1$. However,
in the limit $\beta = \infty$ the GVM gives a continuous
solution which we believe corresponds to a genuine
phase transition.

\subsection{solution for the general case}

We now present the results for the general case of an
elastic manifold in a correlated potential.
Upon rescaling with $\beta$, the solution for the zero mode
$\omega_n=0$ is identical to the problem of uncorrelated disorder
in $d$ dimension. However, the values of the breakpoint $u_c$,
and the quantities associated to it, $[\sigma](u_c) = \Sigma_1$
and $B(u_c)$ are {\it different}. Indeed $\Sigma_1$ and $B(u_c)$ are
determined by equations (\ref{thesame3}): as one sees, the difference
comes from the second equation and $B(u_c)$ is determined by
fluctuations in $d+1$ space rather than $d$. At $T \to 0$
there is no difference and the length $\xi$ is equal
to the Larkin-Ovchinnikov pinning length $R_c$ for the
problem with point disorder in $d$ dimension \cite{giamarchi_vortex_long}.
However, $\Sigma_1$ is
less renormalized by thermal fluctuations and is {\it larger}
than ${\Sigma_1}_{d,uncorr}$ at finite temperature
and thus $\xi < R_c$ at finite
temperature.

Once $\Sigma_1$ is determined, the equations which determine the
connected correlation function, (\ref{fulleq1},\ref{fulleq2})
are closed. The general expression for the correlation function is then:
\begin{equation} \label{funccorgen1}
\tilde{B}(x,\tau) = \overline{\langle[\phi(x,\tau)-\phi(0,0)]^2\rangle}=
\tilde{B}_{d+1}(x,\tau) + \tilde{B}_d(x)
\end{equation}
where the $\tau$ independent part $\tilde{B}_d(x)$ has formally
the same expression as for the
$d$-dim problem with point disorder, with different values
of $\Sigma_1$ and $u_c$. It is the sum of two pieces:

\begin{equation}
\tilde{B}_d(x) = \frac{2 T}{\beta u_c} \int\frac{d^d q}{(2\pi)^d}
                  \left( \frac{\Sigma_1}{c q^2(c q^2+\Sigma_1)}
+ \int_0^1 \frac{dv}{v^2}
\frac{[\sigma](u_c v)}{c q^2(c q^2+[\sigma](u_c v))} (1-\cos(q x)) \right)
\end{equation}
This has a well defined limit when $\beta \to \infty$ since
$\beta u_c$ goes to a constant. We have dropped
the term $\frac{2 T}{\beta} \int \frac{d^d q}{(2\pi)^d}
\frac{1}{c q^2}(1-\cos(q x))$ which corresponds to the
contribution of the zero mode in the absence of disorder,
since it vanishes as $\beta \to \infty$.
The part $\tilde{B}_{d+1}(x,\tau)$ is truly $d+1$ dimensional
and contains the contribution of all the modes. It reads:
\begin{equation}
\tilde{B}_{d+1}(x,\tau)= \frac{2 T}\beta \sum_n \int \frac{d^d q}{(2\pi)^d}
         \frac1{c \omega_n^2+c q^2+\Sigma_1 +
         I(\omega_n)}(1-\cos(q x + \omega_n \tau))
\end{equation}
This contribution has a different scaling than the previous one. The
characteristic
length scale is $\xi = \sqrt{c/\Sigma_1}$ in both $x$ and $\tau$,
and $x \sim \tau^{1/2}$ (diffusion in the localization tube).
$\tilde{B_d}(x)$ has the same long distance behavior as
in the $d$-dim problem and , if the problem has
several length scales, exhibits crossover regimes as
discussed in Ref.~\onlinecite{giamarchi_vortex_long}. The quantities
$I(\omega)$, $B(u_c)$ and
$\Sigma_1$ are determined by equations (\ref{iom0}) and (\ref{thesame3}).

Expanding (\ref{iom0}) to lowest order
in $T$ gives:
\begin{equation} \label{fukuleedd}
I(\omega_n) = 2 \Sigma_1 [ 1 - (1 + \frac{c \omega_n^2
              + I(\omega_n) }{\Sigma_1} )^{(d-2)/2} ]
\end{equation}
after simplifications using the replicon condition
(\ref{thesame3}) which gives
$\Sigma_1 = (- 4 j_d V''(0))^{2/(4-d)}$. (\ref{fukuleedd})
is solved for any $d$ as:
\begin{equation}
I(\omega_n)  =  \Sigma_1 f(\sqrt{c} |\omega_n |/\sqrt{\Sigma_1})
\end{equation}
where $f(z)$ is determined by inverting the equation:
\begin{equation} \label{fukuleesolu}
z^2  = (1 - \frac{(2-d)}{2} f(z))^{2/(d-2)} - f(z) - 1
\end{equation}
For $d \ge 2$ (\ref{fukuleesolu}) is
valid for $\omega_n \ll \Lambda$. The small $z$ expansion
of $f(z)$ is:
\begin{equation} \label{fukuleeseries}
f(z) = \frac{2}{\sqrt{4-d}} z + \frac{2(d-3)}{3(4-d)} z^2 + O(z^{3/2})
\end{equation}
Thus the feature that $I(\omega) \sim |\omega|$ is always verified
(with the choice of the replicon condition) at $T=0$.
One can see easily that the above
low-T solution will give the correct picture of the localized phase.
This is because $\tilde{B}(\tau) - B$ becomes small at large $\tau$.
To obtain the small $\omega$ behavior one can expand
(\ref{iom0}):
\begin{equation} \label{iom3}
I(\omega_n) = V''(B) \frac{2}T \int_0^\beta d\tau
(1-\cos(\omega_n \tau)) (\tilde{B}(\tau) - B) + A(\omega_n)
\end{equation}
where  $A(\omega_n) = \frac{2}T \int_0^\beta d\tau
(1-\cos(\omega_n \tau)) (H(\tilde{B}(\tau)) - H(B))$ and $H(x)=
V(x) - x V'(x)$. The part $A(\omega_n) \sim Z \omega_n^2$ with $Z \sim T$
because the integral:
\begin{equation} \label{iom4}
Z = \frac{2}T \int_0^\infty d\tau \frac{\tau^2}{2} (H(\tilde{B}(\tau)) - H(B))
\end{equation}
is convergent at large $\tau$ as we check self-consistently.
Thus one has:
\begin{equation} \label{fukuleedd2}
I(\omega_n) = 2 \Sigma_1 [ 1 - (1 + \frac{c \omega_n^2
              + I(\omega_n) }{\Sigma_1} )^{(d-2)/2} ] + A(\omega_n)
\end{equation}
where the {\it exact} replicon condition
$\Sigma_1 = (- 4 j_d V''(B))^{2/(4-d)}$
has been used. Expanding in powers of $I(\omega_n)$ and
using $A(\omega_n) \sim Z \omega_n^2$ at
small $\omega_n$, one finds:
\begin{equation}
I(\omega_n) \sim \sqrt{ \frac{4(c+Z) \Sigma_1}{4-d} } |\omega_n|
\end{equation}
This behavior $I(\omega) \sim |\omega|$ at small $\omega$
comes from the decay $B - \tilde{B}(\tau) \sim 1/\tau^2$ at large $\tau$.
This property holds for any temperature inside the
localized phase, as we confirmed
using numerical solution of the saddle point
equations.

\section{Quantum Problems}
\label{section5}

\subsection{models}

The Hamiltonian (\ref{general})
introduced in section~\ref{section2} is also of great
interest for the study of quantum problems. The dimension denoted as $\tau$
is viewed as the time and the Hamiltonian becomes the action of a
quantum problem. Correlated disorder along the ``time'' direction
arises naturally in quantum problems since for these
problems disorder is usually time-independent.
The quantum statistical mechanics of bosons systems
corresponds to considering $\tau$ as the imaginary (Matsubara) time and
taking periodic boundary conditions along the time direction.
The size of the system along the imaginary time direction $\tau$ is,
in this representation, the inverse temperature $\beta$ of the quantum problem.
The temperature $T$ of the corresponding classical problem
is a measure of the quantum
effects, since it corresponds to $\hbar$.
One of the difficulties in treating quantum
problems from the action given in section~\ref{section2} comes from the
fact that all quantities are computed in imaginary time. The physical
dynamical quantities (such as the conductivity) which are of interest for the
quantum problem come from {\it retarded} correlation functions,
obtained from the imaginary time ones via an analytic continuation. Such
a step is in general highly non trivial. Fortunately, as we show here,
it can be performed directly on the variational formulas, allowing to obtain
the physical quantities of interest.

The action (\ref{general}) corresponds to the general quantum problem of a
manifold of internal dimension $d$ ($d=0$ corresponds to a quantum particle)
in a $N$-dimensional random potential. One experimental realization is
the {\it quantum} behavior of flux lines in high-Tc superconductors
($d=1$, $N=2$ for a single line, $d=3$, $N=2$ for an elastic vortex lattice),
which is important in low temperature experiments
\cite{blatter_vortex_review}. The solution given here
for the problem (\ref{general}) is exact in the limit $N \to \infty$.

Many quantum problems can also be described by the
action (\ref{cardy}) and (\ref{cardyos}).
It coincides with the phase Hamiltonian used to describe
charge density waves \cite{fukuyama_pinning,gruner_revue_cdw}
and Wigner crystals
\cite{rusin_shklovskii_wigner,normand_millis_wigner}. The density
of the system depends on a phase $\phi$ by
\begin{equation}
\rho(x) = \rho_0 \cos(Q x + \phi(x))
\end{equation}
where $Q$ is the modulation vector of the charge density wave. The
Hamiltonian describing the energy cost in variation of the phase $\phi$
is
\begin{equation}
{\cal H} = \int d^dx \left( \frac{\Pi^2}{2 M} + (\nabla \phi)^2 \right)
\end{equation}
where $\Pi$ is the conjugate momentum to the phase $\phi$ and $M$ is the
mass of the charge density wave (usually a large number for phonon-induced
charge density waves). If one introduces a random potential
coupling to density and goes to the action representation one obtains
exactly the action (\ref{cardyos}). The current is given by a
continuity equation as $j \sim \partial_t \phi$.

In one dimension (\ref{cardyos}) also describes the problem of
interacting fermions in the presence of disorder. The solution presented here
describes the Anderson localization of interacting particles.
Let us consider for simplicity spinless fermions in
$d=1$ interacting via the
Hamiltonian\cite{solyom_revue_1d,emery_revue_1d,%
shankar_spinless_conductivite}
\begin{equation} \label{hubbard}
{\cal H} = -t\sum_{\langle i,j \rangle} c_{i}^\dagger c_{j} +
     V \sum_i n_{i} n_{i+1}
\end{equation}
where ${\langle i,j \rangle}$ denotes nearest neighbors.
As is well known in one dimension the
fermion operators can be represented in terms of boson
ones\cite{solyom_revue_1d,emery_revue_1d,haldane_bosonisation}.
The bosons represents charge fluctuations of the fermion system.
The complete Hamiltonian (\ref{hubbard})
becomes\cite{solyom_revue_1d,emery_revue_1d,haldane_bosonisation}
\begin{equation} \label{free}
H = \frac1{2\pi}\int dx \left[ (v K)(\pi \Pi)^2
         + (\frac{v}{K}) (\partial_x \phi)^2 \right]
\end{equation}
$\Pi$ and $\phi$ are canonically conjugate variables and $\pi\Pi =
\partial_x\theta$. The constants $v$ (velocity
of excitations) and $K$ incorporate all the effects
of the interaction $V$.
More generally, the Hamiltonian
(\ref{free}) describes correctly all the low energy properties of any
one dimensional spinless fermion system provided one uses the
correct $v$ and $K$ parameters. For the Hamiltonian (\ref{hubbard})
$v$ and $K$ can be computed from the Bethe-Ansatz
solution \cite{luther_spin1/2,haldane_xxzchain},
but we
will consider them in the following as parameters. $K=1$ corresponds
to non-interacting fermions, whereas $K < 1$ (resp. $K > 1$) corresponds
to repulsive (resp. attractive) interaction $V$.
Using (\ref{free}) one gets for the Lagrangian
\begin{equation}  \label{lagdep}
{\cal L} = \frac1{2\pi K} \int dx d\tau [ \frac{1}{v} (\partial_\tau
\phi)^2 + v (\partial_x \phi)^2]
\end{equation}
In the following calculations we will set $v=1$ which can be
done by rescaled time and space.
Therefore in order to identify (\ref{lagdep}) with (\ref{cardyos})
one must have:
\begin{equation} \label{identif}
\frac{c}{2 T} = \frac{1}{2 \pi K}
\end{equation}
The limit $K \to 0$, which is the classical limit of the quantum
problem (at zero temperature $\beta=\infty$),
corresponds to the zero temperature limit $T \to 0$
for the associated classical problem in $d$+1 dimension.
One can now introduce a random potential
\begin{equation}
{\cal L_{\rm imp}} = \sum_i \int d\tau V(i,\tau) n_i(\tau)
\end{equation}
After replicating the fermions and averaging over the random
potential one gets
\begin{equation} \label{fermions}
{\cal L} = -D \int d\tau \sum_{a,b}\sum_i n^a_i(\tau) n^b_i(\tau)
\end{equation}
where $(a,b)$ are the replica indices.
In the continuum limit
the density operator will have Fourier components around
$q=0$ and $q=2k_F$, and using the boson representation
the density becomes
\begin{equation}
n(x,\tau) = \frac1\pi \partial_x \phi(x,\tau) +
          \frac1{2\pi \alpha}
(e^{i 2 k_F x + 2 i\phi(x,\tau)} + {\rm h.c.})
\end{equation}
where $\alpha$ is a cutoff of the order of the lattice spacing.
Physically this means that to describe the low energy properties of the
system one can only consider the Fourier components of the disorder
around $q=0$ and $q=2k_F$, and consider them as independent random
variables. Such an approximation will be valid provided
the disorder is weak enough compared to the Fermi energy,
and therefore if the fermion density is
not too small (for a fixed value of the disorder).
Using the boson representation the disorder term (\ref{fermions})
becomes
\begin{equation} \label{disorder}
{\cal L} = -D \sum_{a \ne b} \int dx d\tau d\tau'
           \frac1{\pi^2}
           (\partial_x \phi^a(x,\tau))(\partial_x \phi^b(x,\tau'))
           + \frac2{(2\pi\alpha)^2}\cos(2\phi^a(x,\tau) -
              2\phi^b(x,\tau'))
\end{equation}
As can be shown either on the fermion or the boson
representation the
$q\sim 0$ part of the disorder can be eliminated by a simple
redefinition of variables, as in (\ref{redef}),
which does not affect the $q \sim 2k_F$
part. Since the current is $j = \frac{1}{\pi} \partial_t \phi$
such a change has no effect on the conductivity, and its effect
on other correlation functions can be simply computed
\cite{giamarchi_loc}. In the
following we therefore keep only the $q\sim 2k_F$ part, which can
lead to the localization of the fermions. The total action we consider
in the following is identical to model
(\ref{cardyos}) with $W/(2 T) = 2 D/(2 \pi \alpha)^2 v$.

A similar mapping can be made for repulsive bosons in one
space dimension
\cite{haldane_bosons,giamarchi_loc}.
The resulting Lagrangian is identical to (\ref{lagdep}) with the
identifications $v/(\pi K) = \rho _0/m$,
and $v^2 = 1/(\kappa \rho _0 m)$. $\kappa$
is the compressibility, $\rho_0$ the average density,
and $m$ the mass of the bosons.
The excited states in the absence of
disorder are sound waves with phase velocity $v$, which are the phonon modes
typical
of a Bose superfluid. The existence of such
modes is sufficient for true superfluidity to
exist \cite{mikeska_supra_1d}.

The point $K=1$ is the best known. It corresponds to non
interacting fermions, or equivalently to bosons with a hard core repulsion.
There, it is well known
that all states are localized by disorder, and that
the conductivity behaves
\cite{berezinskii_conductivity_log,abrikosov_ryzhkin}
as $\sigma(\omega) \sim \omega^2\log^2(\omega)$.
For small disorder one can use the renormalization group (RG)
and one finds a transition between a regime where the disorder is
irrelevant, and a regime where the disorder
renormalizes to large values (there is no perturbative fixed point
accessible in the localized phase). The localized regime is found
to happen for $K < 3/2$, i.e $~T< T_{BG}=3 \pi c/2$ with the identification
(\ref{identif}). This coincides with the value predicted by the
variational method (see section~\ref{sol1d} and Appendix A).
The localization length, the crossover towards the localized regime,
and the high-frequency part of the conductivity
can be also computed using RG \cite{giamarchi_loc}.
When $K\to 0$ the hamiltonian becomes similar to the one
describing the pinning of a charge density wave, for which a self-consistent
formula for the pinning length was obtained by Fukuyama and Lee.
The conductivity was computed analytically for $K=0$ using a transfer
matrix approach and is also found \cite{vinokur_cdw_exact} to behave as
$\sigma(\omega) \sim \omega^2\log^2(\omega)$, as for $K=1$.

Most of the physical properties can be derived from the solution
of the GVM in imaginary time obtained in section~\ref{sol1d}.
Contrarily to the case of the elastic manifold, the length
$\xi$, e.g given by (\ref{pinlength})
(the length at which the phase $\phi$ disorders)
is interpreted as the localization length for
the interacting fermions. This is because it
set the frequency scale in the conductivity
as can be seen on the GVM result below and on the RG.
For $K=1$ it coincides with the usual definition of
the localization length at the Fermi energy.
At a given point in space formula (\ref{locrad})
and (\ref{loclength}) show that
the phase fluctuations saturate to a constant at
large time. Thus, in the localized phase, at each point
there is a well defined average value of the phase that can be viewed as
constant in time but which adjusts spatially in order to take advantage
of the random potential. This agrees with the physical picture
of Fukuyama and Lee \cite{fukuyama_pinning}.

\subsection{conductivity: generalities}

In addition to the results which can be directly transposed
from section~\ref{section4}, let us examine here dynamical correlation
functions. The most
interesting quantity is the conductivity.
We will compute it using the Kubo formula (see Ref.
\onlinecite{mahan_book} Ch. 3
for a derivation, and also Ref.
\onlinecite{shankar_spinless_conductivite}):
\begin{equation} \label{kubo1}
\sigma(q,\omega) =
\frac{i}{\omega}
( \frac{v K}{\pi} + \chi_{ret}(q,\omega) )
\end{equation}
Here $\omega$ is a real frequency associated to real time $t$.
It is understood that $\omega$ always stands for $\omega + i \delta$,
i.e contains an infinitesimal imaginary part. The first term
is the diamagnetic contribution and the second term
is the retarded current-current correlation function:
\begin{equation}
\chi_{ret}(q,\omega) = - i
\int dt e^{i \omega t} \theta(t) \langle
[j^{+}(q,t),j(q,0)] \rangle
\end{equation}
The conductivity is defined as
$\sigma(\omega) = \lim_{q \to 0} \sigma(\omega,q)$. For the
actual calculation one must work with imaginary time $\tau = i t$, use the
imaginary Matsubara frequencies $i \omega_n$ and then perform the
analytic continuation $i \omega_n \to \omega + i \delta$. Introducing:
\begin{equation}
\chi(q, i \omega_n) = - \int_0^{\beta} d\tau e^{i \omega_n \tau}
\langle T_\tau j^{+}(q,\tau) j(q,0) \rangle
\end{equation}
et donc $\chi_{ret}(q,\omega) =
\chi(q,i \omega_n)|_{i\omega_n \to \omega + i\delta}$. Using
$j = \frac1{\pi} \partial_t \phi$,
the integration by part over time in (\ref{kubo1})
cancels the diamagnetic term
\cite{shankar_spinless_conductivite}
and one finally finds the general formula for the conductivity:
\begin{equation} \label{conductivity}
\sigma(\omega) = \frac{1}{\pi^2} \frac{i}{\omega + i \delta} \left(\omega_n^2
\langle T_\tau \phi(q=0,\omega_n)\phi(q=0,-\omega_n) \rangle
            \right)_{i\omega_n \to \omega + i\delta}
\end{equation}

One can apply this formula to a system of interacting spinless fermions
(or bosons) in the absence of disorder. The correlator is then given by
$\langle T_\tau \phi(q,\omega_n)\phi(-q,\omega_n) \rangle
=T/(c \omega_n^2 + c q^2)$ and thus one finds:
\begin{equation} \label{drude}
\sigma(\omega) = \lim_{q \to 0}
\frac{i}{\pi^2(\omega + i \delta)} \left( \frac{T \omega_n^2}{c \omega_n^2 + c
q^2}
\right)_{\omega_n \to - \omega + \delta} =
\frac{i T}{c \pi^2(\omega  + i \delta)} =
D \left( \delta(\omega) + \frac{i}{\pi} P(\frac1{\omega})
\right)
\end{equation}
where P denotes the principal part.
$D= T/(\pi c)= v K$ is the charge stiffness
and the strength of the Drude peak. The charge stiffness
$D$ can be defined quite generally \cite{kohn_stiffness}
by $Re(\sigma(\omega)) =
D \delta(\omega) + \sigma_{reg}(\omega)$. When $D>0$ it means
that the system is a perfect conductor.

To explicit the content of the flux line-quantum boson analogy
it is useful to indicate the relation
between the conductivity of the quantum particles
and the tilt modulus of the corresponding classical
system. Using (\ref{tiltmod}) and (\ref{conductivity})
one has:
\begin{equation}
\left.
\frac{T}{c_{44}(\omega_n)} \right|
_{\omega_n \to - i \omega + \delta}
= - \pi^2 i \omega \sigma(\omega)
\end{equation}
This relation between $c_{44}$, and a dynamic quantity,
$\sigma(\omega)$, involves a non trivial analytic
continuation and one should be careful before
drawing conclusions from one system to the other.
Taking the limit $\omega \to 0$ in the above equation
one gets:
\begin{equation}
\frac{T}{c_{44}}= \lim_{\omega \to 0} \pi^2 Im( \omega
\sigma(\omega) ) = \pi D
\end{equation}
where the last relation follows from the Kramers-Kronig
relations. Thus the tilt
modulus $c_{44}$ is related to the {\it charge stiffness}
$D$ of the quantum system. Both $D$ and $c_{44}$ correspond
to ``dynamical'' quantities, i.e the limit
$q \to 0$ is performed {\it before} the $\omega \to 0$.
Note that both quantities are unrenormalized by interactions
for a galilean invariant system.
In general, $1/ c_{44} \sim D$ should be distinguished
from ``static'' quantities such as
the superfluid density
$\rho_s \sim \partial^2 F/\partial \Phi^2$ which can also
be obtained from the current-current correlation
function, but in the opposite limit
$\omega \to 0$ first then $q \to 0$ (F is the free energy
and $\Phi$ an external flux).
A priori $D$ and $\rho_s$ are different
quantities \cite{giamarchi_shastry_persistent}
(for free fermions $\rho_s=0$
and $D \neq 0$). At $\beta=\infty$,
$D \sim \partial^2 E_0/\partial \Phi^2$ where
$E_0$ is the ground state energy
\cite{kohn_stiffness}.
In $d=2$ and $\beta=\infty$
a relation between $c_{44}$ and $\rho_s$
applies \cite{nelson_columnar_long}
but only when $c_{66}=0$ and thus only in
the liquid phase \cite{hwa_splay_prl,hwa_splay_long},
i.e in the superfluid phase for the bosons.

In the presence of disorder, in the Bose glass phase,
both $\sigma(\omega=0)$ and $c_{44}^{-1}$ are expected to
vanish. A point which deserves further investigation
is whether the two quantities $D$ and $\rho_s$ may generally
become proportional in the presence of
disorder.

\subsection{conductivity computed from the GVM}

In order to compute the conductivity in the most general case
one needs the analytic
continuation of (\ref{fulleq1}-\ref{fulleq2}).
Let us first examine the limit
$K\to 0$, i.e $T\to 0$, where the continuation is
straightforward. The analytical continuation to real frequency $\omega$
can be done directly on the equation (\ref{iom}) and gives:
\begin{equation} \label{iomcont}
I(\omega) = - 4 V''(0) \int \frac{d^dq}{(2\pi)^d} \left(
\frac{1}{c q^2 + \Sigma_1} -
\frac{1}{c q^2 + \Sigma_1 - c \omega^2 + I(\omega)} \right)
\end{equation}
In the case of the single cosine model
(i.e for interacting fermions) in $d=1$ it gives:
\begin{equation} \label{fukuleed20}
I(\omega) = 2 \Sigma_1^{3/2} [ \Sigma_1^{-1/2} - (\Sigma_1 - \omega^2
              + I(\omega))^{-1/2} ]
\end{equation}
The solution of
this equation, and of the general case of
an arbitrary $V(B)$, can be written:
\begin{equation}
I(\omega)  =  \Sigma_1 f(\sqrt{c} i \omega /\sqrt{\Sigma_1})
\end{equation}
where the scaling function $f(z)$ was computed in Section~\ref{section4}.

The expression (\ref{conductivity}) for the conductivity
can be rewritten as:
\begin{equation} \label{conductivity3}
\sigma(\omega) = \frac{ -T \omega}{\pi^2}
\frac{ I''(\omega) + i (I'(\omega) - c \omega^2) }
{I''(\omega)^2 + (I'(\omega) - c \omega^2)^2}
\end{equation}
where $I(i\omega_n)\to I(\omega + i \delta) = I'(\omega) + i I''(\omega)$.
It can also be written in the scaling form:
\begin{equation} \label{conductivity4}
\sigma(\omega) = \frac{ - i T \omega}{\pi^2 \Sigma_1}
s( \sqrt{c} i \omega /\sqrt{\Sigma_1} )
\end{equation}
where $s(z)$ is an analytic scaling function of $z$.
We are primarily interested in the small frequency behavior.
Writing $f(z) = 1 + b z + O(z^2)$, with $b=2/\sqrt{3}$ for
$d=1$, one finds the asymptotic behavior at small frequency:
\begin{eqnarray} \label{result1}
Re( \sigma(\omega) ) & \sim & \frac{ b T \omega^2}{\pi^2 \Sigma_1}
\sqrt{\frac{c}{\Sigma_1}} \nonumber \\
Im( \sigma(\omega) ) & \sim & \frac{ - T \omega}{\pi^2 \Sigma_1}
\end{eqnarray}

To compute the conductivity at positive temperature $T>0$
is more difficult because the analytic continuation
of (\ref{fulleq1}-\ref{fulleq2}) becomes non trivial.
The continuation of the mass term
$\Sigma_1(1-\delta_{n,0})$ give simply $\Sigma_1$, but the
continuation of the self-consistent equation for
$I(i\omega_n)$ is more
involved and is performed
in appendix~\ref{analytic}. The result reads, for
the single cosine model:

\begin{eqnarray} \label{continued}
I^{\text{ret}}(\omega) & = & \frac{4 W e^{-2 B}}{T} \left[
\int_0^\beta d\tau \left(
e^{\frac{4T}{\pi}\int_0^\infty du  A(u)
(e^{- u |\tau|} + 2 N_u \cosh(u \tau))} - 1 \right) \right. \nonumber \\
& & \left. + 2 \int_0^\infty dt e^{i\omega t}
\text{Im} \left[e^{\frac{4T}{\pi}\int_0^\infty du A(u)
(e^{-i u t} + 2 N_u \cos(u t))}
               \right] \right]
\end{eqnarray}
where $N_u = 1/(e^{\beta u} -1)$ is the Bose factor and $A(u)$ is the
$q$ integrated spectral function
\begin{equation}
A(u) = \int \frac{d^dq}{(2\pi)^d}
\frac{I''(u)}{(q^2 - u^2 + \Sigma_1 +
                     I'(u))^2 + (I''(u))^2}
\end{equation}
with the property $A(-u)=-A(u)$.
One must check that indeed in the limit of
$K\to 0$ (equivalently $T\to 0$), when one
can simply expand the exponential one recovers
the above expression (\ref{iomcont}). The expansion
gives:
\begin{equation}
I(\omega) = 16 W e^{-2 B}
\int_0^\infty du A(u) \frac{1}{\pi}
( \frac{2}{u} + \frac{2 u}{(\omega + i \delta)^2 - u^2} )
\end{equation}
Using again (\ref{expo0}) one obtains back (\ref{iomcont}).

We do not expect that finite $T$ ($K>0$) will change
qualitatively the low-frequency result apart from a change in the
coefficient $b$ when $T$ varies, presumably an increase of
$b$ with $T$. The high frequency behaviour however is
strongly dependent on $T$ and is computed
in Appendix~\ref{onedim}.

\subsection{discussion}

We have found a conductivity $\sigma(\omega)
\sim \omega^2$ in any dimension $d$ using the GVM.
This is a direct consequence
of the $i \omega$ term being generated in the self-energy.
While this term appears
naturally when there is full RSB, in $d=1$ for the
single cosine model, it appears
only as the consequence of
the marginality condition in the replica
language. The other choice (minimization
over $u_c$) leads to $I(\omega) \sim \omega^2$ and
to a gap. Indeed generically, simply adding a mass term
to a $\omega^2$ term in the self-energy leads to a gap
in the excitation spectrum.
If one has $\langle T_\tau \phi(q,\omega_n)\phi(-q,\omega_n) \rangle
=T/(c q^2 + c' \omega_n^2 + \Sigma_1)$, then
using (\ref{conductivity}) one finds:

\begin{equation}
\sigma(\omega) \sim
\frac{\omega^2}{\Sigma_1 - \omega^2 - i \delta}
\sim  \delta(\omega- \sqrt{\Sigma_1}) + i P(\frac1{\omega-\Sigma_1})
\end{equation}
Thus there is a gap and no absorption at small frequency.

Within the framework
of the variational method the present result,
$\sigma(\omega) \sim \omega^2$, is obtained by taking into
account only small oscillations around the equilibrium position
(in the following we call ``phonons'' such excitations, not to
be confused with real phonons which are absent here).
Our
result is in reasonable agreement with previous results in $d=1$, up
to logarithmic corrections. Previous results were: (i) for free
fermions at $K=1$, $\sigma(\omega) \sim \omega^2$ by Mott,
and $\sigma(\omega) \sim \omega^2 \log^2 \omega$ through exact
calculation by Berezinskii
\cite{berezinskii_conductivity_log,abrikosov_ryzhkin}
and through a physical
argument by Halperin, (ii) in the classical limit $K=0$,
$\sigma(\omega) \sim \omega^2 \log^2 \omega$ using a transfer matrix
technique \cite{vinokur_cdw_exact}
by Feigelman and Vinokur (FV).
In Mott's physical derivation, the conductivity is estimated
from computing the energy absorbed by the system of electrons.
An external driving frequency $\omega$ will cause electrons
within $\omega$ of the Fermi surface to make transitions
from one localized state to another. The density of the active electrons
is thus $\sim \omega$ and since they absorb a photon of energy $\omega$,
one finally finds a conductivity $\sigma(\omega) \sim \omega^2$.
Thus in this derivation, it seems that the
important excitations are, when cast back in the phase language,
the soliton-like excitations of the phase, corresponding to an
electron being transferred from one localized state to another.
In FV derivation however, only ``phonons'' modes are considered.

One may thus speculate that both solitons and ``phonons'', i.e
small oscillations of
the phase, give the same contribution $\omega^2
\log^2(\omega)$ to the conductivity.
One must note that since we have considered
Gaussian disorder, corresponding to the limit of a large number
of very weak pins, it is always
possible to create phonon-like excitations down to $\omega = 0$.
For Poissonian disorder, on
the other hand, phonon-like excitations
should be impossible \cite{gorkov_cdw_strong}
for frequencies lower than the strength of an
individual pin $V$. In that case the $d=1$ wire is cut
into segments with fixed boundary conditions, and the conductivity
from ``phonons'' is
exponentially activated \cite{gorkov_cdw_strong,fukuyama_pinning}.
On the other hand soliton
like excitations are still allowed. One can therefore expect that
above $\omega = V$ both phonons and solitons will contribute to the
 $\omega^2 \log^2(\omega)$ term in the conductivity, while
below $\omega = V$, only the contribution
of the soliton part should remain. In the
near classical limit, for small $K$,
solitons are energetically much less
favorable than phonons and the contribution of the soliton part should
therefore be small.
The conductivity should therefore remain
as $\omega^2 \log^2(\omega)$ down to arbitrary low temperatures but the
prefactor dropping dramatically below $\omega = V$.
In the other case of large quantum fluctuations,
for example free electrons, there should be practically no
effect of Poissonian disorder.

\section{Conclusions} \label{section6}

In this paper we have extended the application of the
gaussian variational method (GVM) to quantum systems with disorder
and, using the analogy between the two problems,
to classical elastic systems with correlated disorder.
By construction this method is exact in the limit of a
large number of components, i.e in large embedding dimension.
It allows to describe the equilibrium statistical
mechanics of both problems. A glass phase with replica
symmetry breaking (RSB) is found at low temperature.

Compared to the case of point disorder treated in Refs.
\onlinecite{mezard_variational_replica,giamarchi_vortex_long}
the solution in the case of correlated disorder
presents three new non-trivial features.
These are not merely technical complications but
originate from the physics being very different in
the present case.

First a correct treatment of
the boundary conditions is crucial. The disorder being
correlated along $\tau$ (imaginary time for the quantum particles)
a single mode, the mode $\omega=0$, controls most (but not all)
of the physics in the glass phase. The RSB is confined to this mode and
there is a non trivial mechanism to generate an effective mass term
for all the other modes, leading to localization by disorder.
The associated physics which naturally emerges
from the GVM is that of small distortions around a
classical background configuration. The method therefore
confirms the physical picture developed
\cite{fukuyama_pinning} by Fukuyama and
Lee. In addition it allows for a calculation
from first principles of all
correlation functions. The properties of the
classical configuration is fully described by the RSB
and is not assumed from the start. For classical
elastic manifolds we have obtained the pinning length $\xi$, the
localization length $l_\perp$ and the non trivial frequency
dependence of the self-energy $I(\omega)$. The GVM predicts a
space vs ``time'' scaling
$x \sim \tau^z$ with $z=2$ within the localized
phase and up to the transition. For the
single cosine model, this agrees
with a recent RG prediction \cite{balents_loc}
in $d=4 -\epsilon$ obtained directly from the RG fixed point
for the mode $\omega=0$ previously derived in Ref.
\onlinecite{giamarchi_vortex_short,giamarchi_vortex_long}.
These results could in principle be checked experimentally
by measuring correlations along the columns
using neutron diffraction. We have computed the
tilt modulus $c_{44}$ and shown that $c_{44}=\infty$
in the Bose Glass phase, a signature of the transverse Meissner
effect predicted in Ref. \onlinecite{nelson_columnar_long}.
By including higher harmonics and realistic elastic energy
the GVM can be used to obtain a more detailed description
of the Bose Glass phase.

In order to apply the GVM to quantum
problem and compute dynamical
linear response quantities, such as the
conductivity, we had to perform a non trivial
analytic continuation to real time, $t = i \tau$.
We showed that within the GVM such a
continuation can be performed in full generality.
We computed the frequency-dependent conductivity $\sigma(\omega)$
for phase Hamiltonians, relevant e.g. for
Wigner crystals and Charge Density Waves.
These systems are localized by the disorder
and we computed the localization length $\xi$.
We find that $\sigma(\omega) \sim \omega^2$ at low
frequency in any dimension. We have also studied
interacting fermions and bosons in $d=1$. The method
then predicts the localization-delocalization transition correctly.
It yields an analytic form for $\sigma(\omega)$ which
is in agreement with known high
frequency results and which at low frequency
also agrees, up to logarithmic factors,
with the exactly solvable
cases (i.e non-interacting fermions and classical limit).
Such a behavior of the conductivity results from
a new ``friction'' term $i \omega$ generated by the random potential
as predicted by the GVM. Finally, a general relation
between the tilt modulus of the classical problem and the
conductivity of the quantum problem has been derived.

The third new feature arises when looking at the
$d=1$ quantum problem, modeled by a single cosine Hamiltonian.
We found that the natural way to
obtain a physically correct dynamical response
(conductivity) was to use the
so-called marginality condition at the replica
saddle point.
It is known that the marginality condition arises
naturally in the solution \cite{cugliandolo_pspin}
of off-equilibrium Langevin Dynamics (LD) for spin glass models.
The equilibrium
quantum statistical mechanics studied here thus plays the
same role with respect to the full problem of quantum dynamics
as the equilibrium FDT-regime does in the off-equilibrium
Langevin Dynamics (LD) of a classical system
(in which the fluctuation dissipation
theorem (FDT) and time translational invariance holds,
and thus called the FDT-regime).
This scenario agrees with the fact
that the quantum equilibrium dynamical
equations that we obtained (\ref{fulleq1}-\ref{fulleq2})
depend on the quantity $\Sigma_1$ only. This quantity
is determined at the breakpoint from the RS side (short scales)
independently of the details of the RSB at larger scales.
This is completely analogous to the role played by the
``anomaly'' $\overline M$ with respect to the
FDT-regime of the off-equilibrium LD.
The LD for the $d$-dimensional classical
problem (i.e the $\omega_n=0$ mode of the quantum problem)
was studied in Refs.
\onlinecite{cugliandolo_ledoussal_zerod,cugliandolo_ledoussal_manifold}
and the expressions for the anomaly $\overline M$
and for the large time limit $\tilde b_1$ of the displacement
in the FDT-regime were derived.
One can check explicitly that
in the limit $T \to 0$, $\Sigma_1$ and $B$ in the quantum problem
are identical to the anomaly and $\tilde b_1$
from Refs.~
\onlinecite{cugliandolo_ledoussal_manifold,cugliandolo_ledoussal_zerod}.
Thus we speculate that there
also exists an off-equilibrium quantum aging regime,
at finite waiting time $t_w$. The regime studied here
of equilibrium time-translational invariant quantum dynamics
being reached only at infinite waiting time $t_w \to \infty$ and
fixed $\tau=i t$. This aging regime might be relevant
for computing some response functions, in particular
the nonlinear responses.

As we have discussed, the GVM studied here takes into account
only the small gaussian deviations around a non trivial,
spatially dependent, equilibrium (i.e static) configuration.
It would very interesting to extend the method
and take also include kinks and solitons. This
would allow to compute the response to a finite transverse field
(i.e study the transverse Meissner phase)
as well as non linear dynamical quantum responses
and variable range hopping phenomenon.

\acknowledgements

We thank S. Sachdev for very helpful remarks and
M. V. Feigelman, J. Kurchan, B. Spivak and V.M. Vinokur
for interesting discussions.

\appendix

\section{Replica Symmetric Solution}
\label{appendixA}

Assuming replica symmetry in (\ref{eqbase})
one finds:
\begin{eqnarray} \label{eqbase2}
G_c^{-1} & =& c (q^2 + \omega_n^2) + \frac{2}T \int_0^\beta d\tau
               (1-\cos(\omega_n \tau)) [ V'(\tilde{B}(\tau))
                -\int_0^1 du V'(B))] \nonumber \\
\sigma(\omega_n) & = & \frac{2 \beta}{T} \delta_{n,0} V'(B)
\end{eqnarray}
where
\begin{eqnarray}
B & =& \frac{2T}{\beta}\sum_n \int \frac{d^d q}{(2\pi)^d}
G_c(q,\omega_n) \nonumber
\\
\tilde{B}(\tau) & = & \frac{2T}{\beta}\sum_n \int_0^\beta
\frac{d^d q}{(2\pi)^d}
\tilde{G}(q,\omega_n) (1 - \cos(\omega_n \tau))
\end{eqnarray}
Since the $\omega=0$ mode is infrared divergent in $q$,
$B = \infty$ for $d \leq 2$. The equations simplify in:
\begin{eqnarray} \label{eqbase3}
G_c^{-1} & =& c (q^2 + \omega_n^2) + \frac{2}T \int_0^\beta d\tau
               (1-\cos(\omega_n \tau)) V'(\tilde{B}(\tau))
                \nonumber \\
\tilde{B}(\tau) & = & \frac{2T}{\beta}\sum_n \int_0^\beta
\frac{d^d q}{(2\pi)^d}
G_c(q,\omega_n) (1 - \cos(\omega_n \tau))
\end{eqnarray}
Contrarily to the RSB case no good $T=0$ limit exists.
Thus the solution cannot describe a
low temperature localized phase with a finite localization
length. A localized solution of (\ref{eqbase3}) can be found
numerically (adding a mass term), but has no
good limit $\beta \to \infty$ since the integral in
the first equation produces a term proportional
to $\beta$.

The RS solution describes however the high temperature
phase. For power-law models such a phase exists when
in the short-range or marginal case, $\gamma \geq 2/(2-d)$.
For the single cosine model it exists in $d=1$.
It also exists in $d=2$ but since $T_c \propto \beta$
it disappears in the limit $\beta \to \infty$.
Starting from the elastic theory of the flux lattice in $d=2$
correlated disorder always lead to the Bose glass phase. Only the
liquid, such as the case $d=1$ of lines confined to a plane,
may show a transition.

In the high temperature phase, the tilt modulus $c_{44}$ is
renormalized, and given by (in the limit $\beta=\infty$):
\begin{equation}
c_{44} = c + \frac{1}T \int_0^\infty \frac{d\tau}{2 \pi}
\tau^2 V'(\tilde{B}(\tau))
\end{equation}
The divergence of the integral to first order in
perturbation theory signals the transition at $T_{BG}$.
For the model which describes the vortex lattice one
finds $V(x) = - W \exp(-K_0^2 x/2)$, $T_{BG}=6 \pi c/K_0^2$.
Within the self consistent method one has:
\begin{equation}
c_{44} = c + \frac{W K_0^2}T \int_0^\beta \frac{d\tau}{2 \pi}
\tau^2 \exp( - \frac{2 T K_0^2}{\sqrt{4 \pi c_{44} c}} \log(\tau/\Lambda))
\end{equation}
there is no divergence of
$c_{44}$ at the transition, but the solution for $c_{44}$ ceases
to exist below $T_{BG}$. One recovers correctly some of features
of the RG approach.

\section{Study of the RSB solution for the single cosine
model in $d=1$ at finite $T$} \label{onedim}

We now study the equations for the single cosine
model in $d=1$ at finite $T$. We will study the
limit where $\Sigma_1/\Lambda \ll 1$, i.e $\xi \gg a$.
This happens in two cases: (i) at weak disorder $W \to 0$ at fixed
$T$ (ii) near the transition $T \to T_{BG}$. We will
only study the case $\beta=\infty$. We take $c=1$.

The closed system of equation (for the displacements, self energy and
effective mass) is:

\begin{eqnarray}
\hat{B}(\tau) & = & T \int_{-\Omega}^{\Omega} \frac{d \omega}{2 \pi}
\frac{ \cos(\omega \tau) }{\sqrt{\omega^2 + \Sigma_1 + I(\omega)}}
\nonumber \\
I(\omega) & = & \frac{(\Sigma_1)^{3/2}}{T}
\int_0^{\infty} d\tau (1 - \cos(\omega \tau)) (e^{2 \hat{B}(\tau)} -1)
\label{above} \\
(\Sigma_1)^{3/2} & = & 4 W e^{- 2 \hat{B}(0)} \nonumber
\end{eqnarray}
where $\Omega = \Lambda$ is the ultraviolet cutoff, and we have denoted
by $\hat{B}(\tau) = \tilde{B}(\tau) - B$.
All variables can be rescaled when $\Sigma_1 \to 0$.
We thus define $\tau = (\Sigma_1)^{-1/2} u$,
$\omega = (\Sigma_1)^{1/2} w$, and as before
$I(\omega) = \Sigma_1 f(w)$. The cutoff
becomes $w_m = (\Sigma_1)^{-1/2} \Omega$ and we are thus
interested in the limit $w_m \to \infty$.
Eqs. (\ref{above}) become:
\begin{eqnarray}
\hat{B}(u) & = & T \int_{-w_m}^{w_m} \frac{dw}{2 \pi}
\frac{ \cos(w u) }{\sqrt{w^2 + 1 + f(w)}}
\nonumber \\
f(w) & = & \frac{1}{T}
\int_0^{\infty} du (1 - \cos(w u)) (e^{2 \hat{B}(u)} -1)
\nonumber \\
(\Sigma_1)^{3/2} & = & 4 W e^{- 2 \hat{B}(0)} ~~~~ w_m = (\Sigma_1)^{-1/2}
\Omega
\end{eqnarray}
The first two equations
have a well defined limit when $w_m \to \infty$
as long as $T < T_{BG} = 3 \pi/2$. Thus the
functions $\hat{B}(u)$ for $u > 0$ and
$f(w)$ for $w < \infty$ also have a well
defined limit. The approach of
the BG transition from below is characterized by divergences
in the integrals at {\it small} rescaled time $u$.
As we show below, when $T < T_{BG}$ one has $f(w) \ll w^2$ at large
frequency $w \gg 1$. Thus one has
$\hat{B}(u) \sim -(T/\pi) \ln(u/u^{*})$ at small
$u$, and reporting in the second equation one gets:
\begin{equation}
f(w) \sim \frac{1}{T}
\int_0^{\infty} du (\frac{u^{*}}{u})^{\frac{2T}{\pi}} (1 - \cos(w u))
\end{equation}
One sees that this integral converges at small $u$ provided
$T < T_{BG} = 3 \pi/2$.
There is thus a well defined weak disorder continuum limit
in $d=1$.

Analyzing further the solution one finds the following.
For $T< \pi/2$ (i.e $K<1/2$ )
$\lim_{w \to \infty} f(w) = f(\infty)$ is finite. For
$\pi/2 < T < 3 \pi/2$ one has
$f(w) \sim w^{\alpha}$ with $\alpha = \frac{2T}{\pi} - 1$
(for $1 \ll w \ll w_m$. One sees that
the point $K=1$ (free fermions) corresponds
to $f(w) \sim w$ both at small and large frequency. The transition
at $T=T_{BG} = 3 \pi/2$ occurs when $\alpha=2$, i.e
when the term $I(\omega)$ becomes of the same order
as  $\omega^2$ at {\it large frequencies}.
Of course the small frequency behavior
is always the same $\sim |w|$ as discussed in the text.

Using the formula (\ref{conductivity3})
for the conductivity and the above result
$I(\omega) \sim \omega^{2 K - 1}$ at large
$\omega$, one finds:
\begin{equation}
\sigma(\omega) \sim \frac{1}{\omega^{4 - 2K}}
\end{equation}
This power-law behavior of the conductivity coincides with previously
known results \cite{giamarchi_loc,maurey_wigner}.
The GVM therefore describes the correct physical behavior for the
conductivity for all frequency scales.

\section{Remarks on the variational
method} \label{variational}

For finite number of components $N$ the Hartree or Gaussian
Variational Method is only an approximation.
It is important to determine whether it is a
good approximation. It is in principle
possible to answer in a $1/N$ expansion, but this is
a rather difficult task. The GVM seems to work
well e.g for the cosine model in the case of point disorder
where it can be compared with RG calculations.
In $d=4-\epsilon$ the values of the amplitude of logarithmic
growth of displacements was found \cite{giamarchi_vortex_long}
to agree within $10 \%$ in the FRG and GVM. In $d=2$ it predicts the
transition temperature $T_G$ exactly.

In this Appendix we illustrate some of the artifacts
produced by the method. When properly understood these
examples may help decide to which extent and
in which cases the results of the GVM can be trusted.
The following simple model is studied:

\begin{equation}
H = \int d^dx \frac{c}{2} (\nabla \phi)^2  - g \cos{\phi(x)}
\end{equation}
The variational Hamiltonian is
$H_0 = \int d^dq/(2 \pi)^d) (1/2) G^{-1}(q) \phi(q) \phi(-q)$
and the variational free energy
$F_{var} = - T \ln Tr \exp(- H_0/T) +  \langle H-H_0 \rangle_{H_0}$
is found to be:
 \begin{equation}
\frac{F_{var}}{V} = - \frac{T}{2}
\int \frac{d^dq}{(2 \pi)^d}\ln (T G(q))
+ \frac{T c}{2} \int \frac{d^dq}{(2 \pi)^d} q^2 G(q)
- g \exp( - \frac{T}{2} \int \frac{d^dq}{(2 \pi)^d} G(q))
\end{equation}
where $V$ is the volume of the system.
The saddle point equation thus gives
$G^{-1}(q)= c q^2 + \sigma$
with:
\begin{equation}
\sigma = g \exp( - \frac{T}{2} \int \frac{d^dq}{(2 \pi)^d}
\frac{1}{c q^2 + \sigma} )
\end{equation}

Let us consider $d=1$, i.e a string directed along $z$, in a
periodic potential independent of $z$. In the limit of a large
ultraviolet cutoff,
$q_{max}=\Lambda \to \infty$, one finds, up to a
diverging but $\sigma$-independent contribution:
\begin{equation}
\frac{F_{var}}{V} = \frac{T^2}{16 c} ( x - \tilde{g} e^{-1/x} )
\end{equation}
where we have defined $x=4 \sqrt{c \sigma}/T$ and $\tilde{g} = 16 c g/T^2$.
The total field fluctuation is $\langle \phi^2 \rangle =1/2x$.
One sees that at small $g$, $F_{var}$ is a monotonic function
with a minimum at $x=0$. The first term, the entropy,
dominates and the
string is delocalized, $ \langle \phi^2 \rangle=\infty$,
which is the correct result. However,
for $\tilde{g} > e^2/4$ a second minimum at finite
$x$ appears (it appears at $x=1/2$ for $\tilde{g} = e^2/4)$.
There are then three roots to the saddle point
equation. The absolute minimum remains in
$x=0$ until $\tilde{g}$ reaches $\tilde{g}=e$ (when $x=1$),
at which the (unphysical) finite $x$ solution becomes of lower free
energy.

The fact that in $d=1$ the GVM predicts a transition
at $\tilde{g}=e$ to a localized phase is
an artifact. The physical reason for this
result is very clear, however.
The first term in the variational free energy is
the entropy cost $\sim T^2/ c R^2$ of confining the
string to a transverse extension $R$, i.e the kinetic energy of
a bound state in the quantum mechanical analogy. This is a correct estimate.
The energy cost of delocalizing the string away from
a minimum is, however,
strongly {\it overestimated} at large $\tilde{g}$. Indeed
it is computed by assuming that the localized packet
is gaussian, while in fact the packet will be strongly peaked near
the minima of the potential.
The true delocalization will thus proceed via kinks and
be less costly in energy. At small
$\tilde{g}$ this effect is not a problem, and one expects
the method to work well.

In the case $d=0$ a similar artifact occurs, though in a
weaker sense. The equation $ \sigma = g \exp( - T/(2 \sigma))$
has always the physically correct solution
$ \sigma = 0$. If $g/T > e/2 $
two additional solutions appear: $ \sigma = \sigma_{+}$
and $ \sigma = \sigma_{-}$, such that
at $g/T= e/2 $,  $\sigma_{-}= \sigma_{+} = T/2$.
However in the GVM in $d=0$ entropy always wins
and the absolute minimum in the variational free
energy is at
$\langle \phi^2 \rangle =\infty$, which is the correct physical result.
At large enough $g$, however, a second unphysical minimum
appears, which leads to some artifacts
in the presence of an extra mass term.

Our conclusion is thus that one may trust the GVM only
for weak coupling or disorder. A transition at intermediate
couplings, in low dimension, which cannot be tuned
to zero, is usually the sign of an artifact. We believe
a similar artifact to occur in the case of $d=1$
with point disorder (see Appendix C in
Ref. \onlinecite{giamarchi_vortex_long}
and Section~\ref{sol1d} above). This artifact may be indicative of
tendencies of the system, e.g. to be more glassy, but not taken
as the sign of a true transition. In the disordered case,
one may also be able to determine when
statistical entropy and disorder energy are incorrectly evaluated.
One generally expects that the main corrections should be
produced by kinks or instanton configurations, while
small displacements, e.g. phonons, should be described correctly.

\section{Analytic continuation} \label{analytic}

We illustrate the method of
analytic continuation to real time on the
single cosine model (\ref{cardyos}),
with $V(x) = - W \exp(-2 x)$, which is relevant for
interacting fermions in $d=1$. The extension
to arbitrary $V(x)$ is straightforward.

We want to perform the analytical continuation of:

\begin{equation} \label{depart}
F(i\omega_n) = \frac{4 W e^{-2 B}}{T} \int_0^\beta d\tau
(1 - \cos(\omega_n \tau))
\left(e^{\frac{4T}{\beta} \int \frac{dq}{2\pi} \sum_p
         \frac{\cos(\omega_p \tau)}{c(\omega_p^2 + q^2) + \Sigma_1 +
          I(i\omega_p)}} - 1 \right)
\end{equation}
where $I(i \omega_n)$ is
the self energy with Matsubara (imaginary)
frequencies. Physical quantities are expressed in terms of the
retarded self-energy
$I(\omega) = I'(\omega) + i I''(\omega)$ obtained
by the continuation $i \omega_n \to \omega + i \delta$.
The term coming from the $1$ in the integral (\ref{depart}) is a
simple constant term,
independent of frequency and is therefore unchanged under the analytic
continuation. Difficulties come from the cosine term. This term is the
Fourier transform in Matsubara frequencies of the imaginary time
function
\begin{equation} \label{imagtime}
I_{\text{cos}}(\tau) =
 e^{\frac{4T}{\beta} \int \frac{d^dq}{(2\pi)^d} \sum_p
         \frac{\cos(\omega_p \tau)}{c(\omega_p^2 + q^2) + \Sigma_1 +
          I(i\omega_p)}} - 1
\end{equation}
since the $\omega_n = 2\pi n/\beta$ are Matsubara boson frequencies.
In general, obtaining $I^{\rm ret}(t)$ from (\ref{imagtime}), is
a difficult task. Here, however, $I$ is computed using the Gaussian
variational approximation. Within this approximation
$I_{\text{cos}}$ is {\it itself} a Green function. One can therefore use
for $I_{\text{cos}}$
the well known relation between the real time time ordered boson
Green function
\begin{equation} \label{simple1}
I_{\text{cos}}(t) = \theta(t) I_1(t) + \theta(-t) I_2(t)
\end{equation}
and the retarded one to get
\begin{equation}\label{simple2}
I^{\rm ret}_{\text{cos}}(t) = \theta(t) [I_1(t) - I_2(t)]
\end{equation}
to perform the analytical continuation, it is therefore sufficient to
know the real time, time ordered, function $I_{\text{cos}}(t)$.

In order to do so, we use a spectral representation.
\begin{equation} \label{expo0}
G_c[q,i \omega_p] =
\frac{1}{c(\omega_p^2 + q^2) + \Sigma_1 +
I(i\omega_p)} = \frac{-1}{\pi} \int_{-\infty}^\infty
du A(q,u) \frac{1}{i \omega_p - u}
\end{equation}
where the spectral function $A(q,u)$ is the imaginary part of the
retarded function (see Ref. \onlinecite{mahan_book} sections 3-2 to 3-5)
defined by
\begin{equation}
A(q,u) = Im (G_c[q,i \omega_p \to \omega + i \delta])
=\frac{I''(u)}{(c(q^2 - u^2) + \Sigma_1 +
                     I'(u))^2 + (I''(u))^2}
\end{equation}
The above identity uses $\delta(\omega-u) = \frac{-1}{\pi} Im[
1/(\omega + i\delta - u)]$. The real part of
(\ref{expo0}) gives $Re(G_c[q,i \omega_p \to \omega + i \delta])
=\frac{-1}{\pi} \int du A(q,u) P(1/(\omega -u))$, which
is the Kramers-Kronig relation.
The term in the exponential in (\ref{imagtime}) then becomes
\begin{equation} \label{expo}
\frac{-4T}{\beta\pi} \int \frac{d^dq}{(2\pi)^d}\int_{-\infty}^\infty
du \sum_p A(q,u) \frac{\cos(\omega_p \tau)}{i \omega_p - u}
\end{equation}
Using the fact that $A(q,-u) = -A(q,u)$ for bosons systems
one can rewrite (\ref{expo}) as
\begin{equation} \label{phonon}
\frac{4T}{\beta\pi} \int \frac{d^dq}{(2\pi)^d}\int_0^\infty du
\sum_p \cos(\omega_p \tau) \frac{2 u}{\omega_p^2 + u^2} A(q,u)
\end{equation}
One recognize in (\ref{phonon}) the Fourier transform of a simple
phonon propagator which gives
\begin{equation}
\frac{4T}{\pi} \int \frac{d^dq}{(2\pi)^d} \int_0^\infty du
[e^{-|\tau| u} + 2 N_u \cosh(u \tau)] A(q,u)
\end{equation}
where $N_u= 1/(e^{\beta u} -1)$ is the boson occupation factor.
The time continuation to real time $\tau=i t$ is now
easily performed \cite{mahan_book} and gives
\begin{equation}
\frac{4T}{\pi} \int \frac{d^dq}{(2\pi)^d}\int_0^\infty du
[e^{-i |t| u} + 2 N_u \cos(u t)] A(q,u)
\end{equation}
Using (\ref{simple1}) and (\ref{simple2}) one finally obtains
\begin{eqnarray}
I^{\text{ret}}(\omega) & = & \frac{4 W e^{-2 B}}{T} \left[
\int_0^\beta d\tau \left(
e^{\frac{4T}{\pi}\int_0^\infty du  A(u)
(e^{- u |\tau|} + 2 N_u \cosh(u \tau))} - 1 \right) \right. \nonumber \\
& & \left. + 2 \int_0^\infty dt e^{i\omega t}
\text{Im} \left[e^{\frac{4T}{\pi}\int_0^\infty du A(u)
(e^{-i u t} + 2 N_u \cos(u t))}
               \right] \right]
\end{eqnarray}
where $A(u) = \int\frac{d^dq}{(2\pi)^d} A(q,u) $ is the $q$ integrated
spectral function.

%\bibliographystyle{prsty}
%\bibliography{revues,total,column}

\begin{thebibliography}{10}

\bibitem[*]{junk}
Laboratoire Associ\'e au CNRS. email: giam@lps.u-psud.fr.

\bibitem[\dag]{frad}
Laboratoire Propre du CNRS, associ\'e \'a l'Ecole Normale Sup\'erieure et \`a
  l'Universit\'e Paris-Sud. email: ledou@physique.ens.fr.

\bibitem{blatter_vortex_review}
G. Blatter {\it et~al.}, Rev. Mod. Phys. {\bf 66},  1125  (1994).

\bibitem{civale_columnar_first}
L. Civale and {et al.}, Phys. Rev. Lett. {\bf 67},  648  (1991).

\bibitem{konczykowski_columnar_first}
M. Konczykowski and {et al.}, Phys. Rev. B {\bf 44},  7167  (1991).

\bibitem{nelson_columnar_long}
D.~R. Nelson and V.~M. Vinokur, Phys. Rev. B {\bf 48},  13060  (1993).

\bibitem{hwa_splay_prl}
T. Hwa, P.~L. Doussal, D.~R. Nelson, and V.~M. Vinokur, Phys. Rev. Lett. {\bf
  71},  3545  (1993).

\bibitem{ledoussal_nelson_splay}
P. {Le Doussal} and D.~R. Nelson, Physica C {\bf 232},  69  (1994).

\bibitem{gruner_revue_cdw}
G. Gr{\" u}ner, Rev. Mod. Phys. {\bf 60},  1129  (1988).

\bibitem{rusin_shklovskii_wigner}
I.~M. Rusin, S. Marianer, and B.~I. Shklovskii, Phys. Rev. B {\bf 46},  3999
  (1992).

\bibitem{normand_millis_wigner}
B.~G.~A. Normand, P.~B. Littlewood, and A.~J. Millis, Phys. Rev. B {\bf 46},
  3920  (1992).

\bibitem{sorensen_bosons_disorder}
E.~S. Sorensen, M. Wallin, S.~M. Girvin, and A.~P. Young, Phys. Rev. Lett. {\bf
  69},  828  (1992).

\bibitem{feynman_hibbs}
R.P. Feynman and A.R. Hibbs, {\it Path Integrals and Quantum Mechanics}
  (Mac-Graw Hill, New York, 1965); R. P. Feynman, {\it Statistical Mechanics}
  (Benjamin, Reading, MA, 1972).

\bibitem{nelson_seung_long}
D.~R. Nelson and S. Seung, Phys. Rev. B {\bf 39},  9153  (1989).

\bibitem{nelson_ledoussal_liquid}
D.~R. Nelson and P. {Le Doussal}, Phys. Rev. B {\bf 42},  10113  (1990).

\bibitem{boson_mapping}
E.L. Pollock and D.M. Ceperley Phys. Rev. B {\bf 36} 8343 (1987); Teitel.

\bibitem{kamien_ledoussal_nelson}
R. Kamien, P. {Le Doussal}, and D.~R. Nelson, Phys. Rev. A {\bf 45},  8727
  (1992).

\bibitem{chen_teitel_prl1}
T. Chen and S. Teitel, Phys. Rev. Lett. {\bf 74},  2792  (1995).

\bibitem{li_teitel_prb1}
Y.~H. Li and S. Teitel, Phys. Rev. B {\bf 49},  4136  (1994).

\bibitem{fisher_bosons_scaling}
M.~P.~A. Fisher, P.~B. Weichman, G. Grinstein, and D.~S. Fisher, Phys. Rev. B
  {\bf 40},  546  (1989).

\bibitem{wallin_girvin_bosons}
M. Wallin and S.~M. Girvin, Phys. Rev. B {\bf 47},  14642  (1993).

\bibitem{krauth_bosons_disorder}
W. Krauth, N. Trivedi, and D. Ceperley, Phys. Rev. Lett. {\bf 67},  2307
  (1991).

\bibitem{makivic_trivedi_bosons}
M. Makivic, N. Trivedi, and S. Ullah, Phys. Rev. Lett. {\bf 71},  2307  (1993).

\bibitem{batrouni_hwa}
For recent simulations in $d=1$ see G. Batrouni and T. Hwa, to be published.

\bibitem{giamarchi_loc}
T. Giamarchi and H.~J. Schulz, Phys. Rev. B {\bf 37},  325  (1988).

\bibitem{berezinskii_conductivity_log}
V.~L. Berezinskii, Sov. Phys. JETP {\bf 38},  620  (1974).

\bibitem{mezard_variational_replica}
M. Mezard and G. Parisi, J. de Phys. I {\bf 1},  809  (1991).

\bibitem{hwa_fisher_flux}
T. Hwa and D.~S. Fisher, Phys. Rev. Lett. {\bf 72},  2466  (1994).

\bibitem{hwa_fisher_long}
T. Hwa and D. Fisher, Phys. Rev. B {\bf 49},  3136  (1994).

\bibitem{giamarchi_vortex_short}
T. Giamarchi and P. {Le Doussal}, Phys. Rev. Lett. {\bf 72},  1530  (1994).

\bibitem{giamarchi_vortex_long}
T. Giamarchi and P. {Le Doussal}, Phys. Rev. B {\bf 52},  1242  (1995).

\bibitem{korshunov_variational_short}
S.~E. Korshunov, Phys. Rev. B {\bf 48},  3969  (1993).

\bibitem{nattermann_pinning}
T. Nattermann, Phys. Rev. Lett. {\bf 64},  2454  (1990).

\bibitem{villain_cosine_realrg}
J. Villain and J.~F. Fernandez, Z Phys. B {\bf 54},  139  (1984).

\bibitem{ledoussal_rsb_prl}
P. {Le Doussal} and T. Giamarchi, Phys. Rev. Lett. {\bf 74},  606  (1995).

\bibitem{giamarchi_columnar_short}
T. Giamarchi and P. {Le Doussal}, Physica C {\bf 235-240},  2683  (1994).

\bibitem{fukuyama_pinning}
H. Fukuyama and P.~A. Lee, Phys. Rev. B {\bf 17},  535  (1978).

\bibitem{suzumura_scha}
Y. Suzumura and H. Fukuyama, J. Phys. Soc. Jpn. {\bf 52},  2870  (1983).

\bibitem{giamarchi_vortex_comment}
{\it Translational order and neutron diffraction studies of magnetic flux
  lattices}, T. Giamarchi, P. Le Doussal, submitted to Phys. Rev. Lett.

\bibitem{subir_replicas}
S. Sachdev, Private Communication.

\bibitem{bouchaud_variational_vortex}
J.~P. Bouchaud, M. M{\'e}zard, and J.~S. Yedidia, Phys. Rev. B {\bf 46},  14686
   (1992).

\bibitem{solyom_revue_1d}
J. S\'olyom, Adv. Phys. {\bf 28},  209  (1979).

\bibitem{emery_revue_1d}
V.~J. Emery,  in {\em Highly Conducting One-Dimensional Solids}, edited by
  J.~T.~D. et~al. (Plenum, New York, 1979), p.\ 327.

\bibitem{shankar_spinless_conductivite}
R. Shankar, Int. Review of Modern Physics B {\bf 4},  2371  (1990).

\bibitem{haldane_bosonisation}
F.~D.~M. Haldane, J. Phys. C {\bf 14},  2585  (1981).

\bibitem{luther_spin1/2}
A. Luther and I. Peschel, Phys. Rev. B {\bf 12},  3908  (1975).

\bibitem{haldane_xxzchain}
F.~D.~M. Haldane, Phys. Rev. Lett. {\bf 45},  1358  (1980).

\bibitem{haldane_bosons}
F.~D.~M. Haldane, Phys. Rev. Lett. {\bf 47},  1840  (1981).

\bibitem{mikeska_supra_1d}
H.~J. Mikeska and H. Schmidt, J. Low Temp. Phys {\bf 2},  371  (1970).

\bibitem{abrikosov_ryzhkin}
A.~A. Abrikosov and J.~A. Ryzhkin, Adv. Phys. {\bf 27},  147  (1978).

\bibitem{vinokur_cdw_exact}
M.~V. Feigelmann and V.~M. Vinokur, Phys. Lett. A {\bf 87},  53  (1981).

\bibitem{mahan_book}
G.~D. Mahan, {\em Many Particle Physics} (Plenum, New York, 1981).

\bibitem{kohn_stiffness}
W. Kohn, Phys. Rev. {\bf 133},  A171  (1964).

\bibitem{giamarchi_shastry_persistent}
T. Giamarchi and B.~S. Shastry, Phys. Rev. B {\bf 51},  10915  (1995).

\bibitem{hwa_splay_long}
T. Hwa, P. Le Doussal, D. R. Nelson, V. M. Vinokur, to be published.

\bibitem{gorkov_cdw_strong}
L.~P. Gorkov, JETP Lett. {\bf 25},  358  (1977).

\bibitem{balents_loc}
L. Balents, Europhys. Lett. {\bf 24},  489  (1993).

\bibitem{cugliandolo_pspin}
L. F. Cugliandolo and J. Kurchan; Phys. Rev. Lett. {\bf 71}, 173 (1993); Phil.
  Mag. {\bf B71} (1995).

\bibitem{cugliandolo_ledoussal_zerod}
L. F. Cugliandolo and P. {Le Doussal}, preprint cond-mat/9505112, to appear in
  Phys. Rev. E (1995).

\bibitem{cugliandolo_ledoussal_manifold}
L. F. Cugliandolo,  J. Kurchan and P. {Le Doussal}, to be published.

\bibitem{maurey_wigner}
H. Maurey and T. Giamarchi, Phys. Rev. B {\bf 51},  10833  (1995).

\end{thebibliography}

\end{document}